\documentclass{jfm}

\usepackage{makecell}
\usepackage{caption}
\usepackage{comment}
\usepackage[outdir=./]{epstopdf}
\usepackage{newtxtext}
\usepackage{newtxmath}
\usepackage{natbib}
\usepackage{hyperref}
\hypersetup{
    colorlinks = true,
    urlcolor   = blue,
    citecolor  = blue,
}

\newcommand{\RomanNumeralCaps}[1]
\linenumbers

\title{Stability of acoustic streaming jets}

\author{Bjarne Vincent\aff{1,2},
  Abhishek Kumar\aff{2},
  Daniel Henry\aff{1},
  Sophie Miralles\aff{1}\corresp{\email{sophie.miralles@insa-lyon.fr}},
  Valéry Botton\aff{1},
  \and Alban Pothérat\aff{2}\corresp{\email{alban.potherat@coventry.ac.uk}}}

\affiliation{\aff{1}INSA Lyon, CNRS, Ecole Centrale de Lyon, Universite Claude Bernard Lyon 1, Laboratoire de Mécanique des Fluides et d'Acoustique, UMR5509, 69621, Villeurbanne France
\aff{2}Fluid and Complex Systems Research Centre, Coventry University, Coventry CV1 5FB, UK}

\begin{document}
\maketitle

\begin{abstract}
We study the stability of a steady Eckart streaming jet flowing in a closed cylindrical cavity. This configuration is a generic representation of industrial processes where driving flows in cavity by means of acoustic forcing offers a contactless of stirring or controlling confined flows. Successfully doing so however requires sufficient insight into the topology induced by the acoustic beam. This, in turn, raises the more fundamental question of whether the basic jet topology is stable and, when it is not, of the alternative states that end up being acoustically forced. To answer these questions we consider a flow forced by an axisymmetric diffracting beam of attenuated sound waves emitted by a plane circular transducer at one cavity end. At the opposite end, the jet impingement drives recirculating structures spanning nearly the entire cavity radius. We rely on Linear Stability Analysis (LSA) together with three-dimensional nonlinear simulations to identify the flow destabilisation mechanisms and to determine the bifurcation criticalities. We show that flow destabilisation is closely related to the impingement-driven recirculating structures, and that the ratio $C_R$ between the cavity and the maximum beam radii plays a key role on the flow stability. In total, we identified four mode types destabilising the flow. For $4 \leq C_R \leq 6$, a non-oscillatory perturbation rooted in the jet impingement triggers a supercritical bifurcation. For $C_R = 3$, the flow destabilises through a subcritical non-oscillatory bifurcation, and we explain the topological change of the unstable perturbation by analysing its critical points. Further reducing $C_R$ increases the shear within the flow, and gradually moves the instability origin to the shear layer between impingement-induced vortices: for $C_R = 2$, an unstable travelling wave grows out of a subcritical bifurcation, which becomes supercritical for $C_R=1$. For each geometry, the nonlinear 3D simulations confirm both the topology and the growth rate of the unstable perturbation returned by LSA. This study offers fundamental insight into the stability of acoustically-driven flows in general but also opens possible pathways to either induce turbulence acoustically, or to avoid it in realistic configurations.
\end{abstract}


{\bf MSC Codes }  {\it(Optional)} Please enter your MSC Codes here

\section{Introduction}
\label{sec:introduction}

This work is concerned with the stability of a jet flow driven by an axisymmetric beam of attenuated travelling sound waves in a closed cavity. We focus here on the effect of confinement on the stability of the jet, and shed light on the very unstable nature of such flows.

Sound-driven flows are referred to as acoustic streaming, and result from a nonlinear process producing momentum out of the attenuation of sound waves. Acoustic streaming can be divided into two categories, depending on whether the flow is forced by either standing (Rayleigh-Schlichting streaming \citep{Rayleigh1884,Westervelt1953}) or travelling (Eckart streaming \citep{Eckart1948}) sound waves. Since this work focuses exclusively on flows driven by travelling sound waves, any mention of acoustic streaming from now shall refer to Eckart streaming, unless specified otherwise.

As they allow for remote flow creation, both Rayleigh and Eckart streaming recently received a renewed interest in fields for which geometrical constraints prevent fluid actuation by mechanical means, e.g., using a propeller. This is typically the case in microfluidics, where Rayleigh and Eckart streaming are commonly used to promote flow in microcavities and microchannels \citep{Hagsater2007,Frommelt2008,Friend2011,Muller2015}, in droplets \citep{Alghane2012,Riaud2017}, and around oscillating bubbles \citep{Cleve2019,Doinikov2022,Fauconnier2022}.

At a larger scale, acoustic streaming offers an advantageous mean of actuating fluids whilst avoiding the contamination inherent to mechanical contact. This situation commonly arises in metals and semiconductors solidification processes, during which flows in the melt may affect the quality of the solid ingot. By stabilising convective flows within the melt~\citep{Dridi2008a}, acoustic streaming is known to enhance the homogenity of the solid ingots~\citep{Kozhemyakin2003}. Solidification processes may also benefit from a local control of the grain growth obtained by directing a streaming jet towards the solidification front~\citep{Lebon2019a}. More recently, \citet{Absar2017} showed that nanocomposites manufacturing can also benefit from acoustic streaming to homogenise the distribution of fibers and particles within the metal matrix. All these applications of acoustic streaming to materials manufacturing rely on low-frequency and high-power ultrasounds (typically a few kilohertz and hundreds of watts) which, close to the source, create a cloud of cavitation bubbles acting as the main sound attenuation mechanism driving the flow. Although this setting leads to strong jets, it suffers from the loss of beam coherence at short distances from the source. This prevents any remote control of the flow. We are instead interested in using beams of low-power and megahertz sound waves to extend the range over which controlled flow actuation is possible~\citep{Kamakura1996,Mitome1998}. By acting deep into the bulk of the flow, such a choice of parameters may open new possibilities of usingbacoustic streaming in solidification processes. Examples include the local enhancement of mass transfer at the solidification front with an elongated jet~\citep{ElGhani2021}, and the creation of three-dimensional flows using multiple reflections of a single beam to stir the melt~\citep{Vincent2024}. Still, whether the purpose is to drive a complex chaotic flow~\citep{Cambonie2017,Launay2019,Qu2022} or to stabilise the melt~\citep{Dridi2008a}, these applications require a detailed understanding of the flow patterns for a given configuration and beam intensity. Hydrodynamic instabilities may lead to the formation of flow patterns with a topology that significantly differs from that of the forcing beam and even to turbulence.  Whether such states are desired or not, understanding how a streaming jet may destabilise is key to designing optimal industrial setups involving acoustic streaming.

The destabilisation of Eckart streaming jets has been observed in several experiments, such as in those of \citet{Green2016}. The authors studied a streaming flow in a vertical cylindrical cavity filled with a mixture of diethyl phtalate and ethanol. The circular transducer, placed at the top endwall of the cavity, fired bursts of ultrasonic waves travelling towards the opposite end of the fluid volume, where a layer of sound-absorbing material prevented the backward reflection of the incident sound waves. This setting created a steady streaming flow, which however transitioned towards another state after some time due to a thermal instability arising from the sound-absorbing plate progressively heating up. \citet{Moudjed2014b} also reported strong distortion of the streaming jet when increasing the forcing magnitude. These severe and intermittent jet oscillations systematically occurred near the wall facing the transducer, suggesting that the impingement plays a major role on the flow stability.

These experimental observations of acoustic streaming destabilisation motivated theoretical stability analyses. The first acoustic streaming stability work traces back to~\Citet{Dridi2010}, who considered an analytical 1D-1C (one-dimensional, one-component) jet velocity profile in a 2D infinite layer. For various ratios of the source height to the layer thickness, the authors showed that the flow is destabilised by a wave travelling in the direction facing the acoustic forcing. The frequency of that wave decreases as the layer height is increased with respect to the source height. Later, multidimensional simulations in a closed cavity revealed that the destabilisation is indeed rooted in the jet impingement~\citep{BenHadid2012}, as observed by \citep{Moudjed2014b}. Since then, numerous studies focused on how acoustic streaming may stabilise flows driven by other phenomena such as lateral~\citep{Dridi2008} and vertical (i.e., Rayleigh-Bénard) heating~\citep{Henry2022}, gradient of chemical concentration~\citep{Lyubimova2019} and thermodiffusion~\citep{Charrier-Mojtabi2012,Charrier-Mojtabi2019}.  All these studies, however, suffer from a rather crude modelling of the acoustic field driving the flow. First, none of these works considered beam diffraction. The radial enlargement of the beam not only prescribes a radial length scale to the jet~\citep{Moudjed2014,Vincent2024scaling}, but also locally reduces the magnitude of the acoustic forcing due to the conservation of acoustic power along the beam. As such, these two effects create longitudinal gradients of the streamwise jet velocity~\citep{Kamakura1996,Mitome1998,Dentry2014} that cannot be recovered in a non-diffracting approximation. Second, none of the former stability studies accounted for the decay of the forcing along the beam caused by sound attenuation. This approximation violates two fundamental properties of streaming jets: First, attenuation causes the streamwise jet momentum flux to saturate at infinite distances from the source. Second,  attenuation defines the length scale over which the jet builds up momentum~\citep{Lighthill1978}. As a consequence, this non-attenuated approximation actually severely overestimates the jet velocity, and fails to capture the jet's structure. Therefore, the causes of destabilisation of realistic jets forced by a diffracting beam of attenuated sound waves still remains an open question.

The aim of this work is precisely to determine how a laminar streaming jet driven by an attenuated and diffracting ultrasound beam may become unstable. We shall consider the simplest geometry for this purpose, i.e. a closed cylindrical cavity with impermeable and sound-absorbing boundaries. The acoustic source, placed at one cavity end, radiates an attenuated beam driving a swirl-free axisymmetric jet. The key novelty is to use a realistic model for the jet accounting for both attenuation and diffraction based on our recent work \citep{Vincent2024scaling}. With these models for the cavity and acoustic beam, the problem loses streamwise invariance. The base flow must be determined by numerical simulations and its bi-global stability is carried out numerically too. To assess the influence of confinement, we consider different cavity sizes based on the attenuation length of the sound waves and on the maximum beam diameter. We address the problem by means of numerical methods based on high-order spectral elements to answer the following questions:
\begin{enumerate}
    \item What are the mechanisms underpinning the flow's instability and what are the conditions of their onset?
    \item What is the type (oscillatory or not) of the unstable perturbation, and how does its topology change when varying the cavity size with respect to the size of the forcing field?
    \item What is the nature of the bifurcations leading to these modes, and in particular, are they subcritical or supercritical?
    \item What are the nonlinear states arising out of the instability?
\end{enumerate}
This work is organised as follows: we shall first define the problem along with the governing equations in \S~\ref{sec:problem_definition}, before presenting in \S~\ref{sec:computational_methodology} the computational methods together with the grid convergence analyses in \S~\ref{sec:computational_methodology}. We then analyse the effect of confinement on the flow structure in \S~\ref{sec:base_flow}, before turning to the LSA in \S~\ref{sec:stability_analysis}. Then, we shall determine the bifurcation criticalities for all cavity sizes in \S~\ref{sec:characterisation_of_the_bifurcations}. Concluding remarks are given in \S~\ref{sec:conclusion}.

\section{Formulating the stability problem for jets driven by attenuated and diffracting progressive waves}
\label{sec:problem_definition}

\subsection{Studied configuration and governing equations}
\label{subsec:studied_configuration}

\begin{figure}
    \centering
    \includegraphics[scale=0.35]{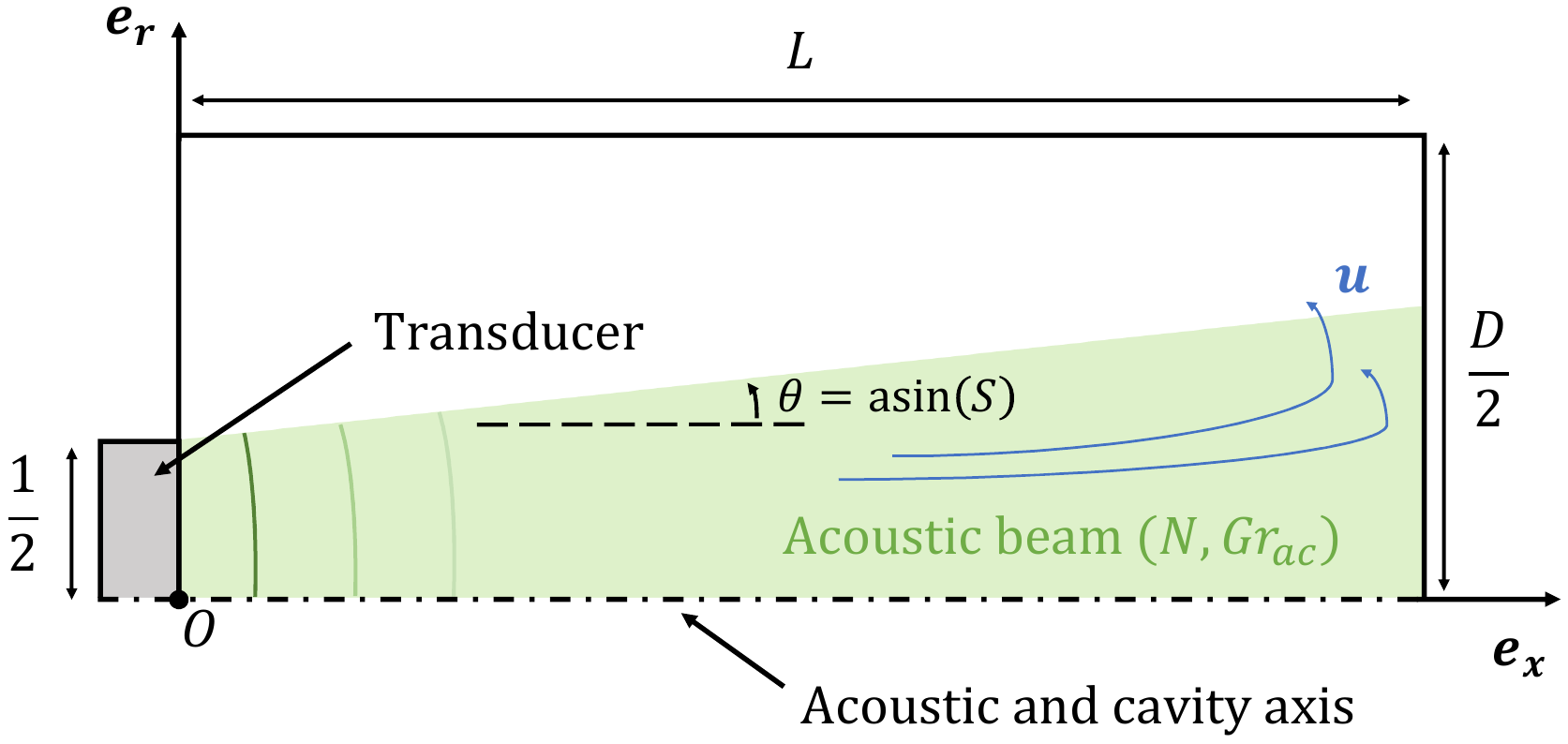}
    \caption{Sketch of the investigated setup in the $\left( x, r \right)$ plane, where $x$ and $r$ are the axial and radial coordinates, respectively. The circular transducer of unit diameter (grey rectangle) at $x=0$ emits a beam of linear sound waves (green shaded area) within an elongated cylindrical cavity of length $L$ and diameter $D$. As the acoustic pressure waves travel across the cavity, their amplitude decay at a rate $N/L$, where $N$ is the ratio between the cavity length and the acoustic pressure attenuation length. Attenuation of these waves yields a body force driving the streaming flow of velocity $\boldsymbol{u}$. All the boundaries are impermeable walls and completely absorb the acoustic waves.}
    \label{fig:sketch_domain}
\end{figure}

We consider an axisymmetric acoustic streaming jet flowing in a closed cylindrical cavity (figure~\ref{fig:sketch_domain}).
The problem is defined using a cylindrical coordinate system in which $\boldsymbol{e_x}$, $\boldsymbol{e_r}$  and $\boldsymbol{e_\theta}$ refer to the axial, radial and azimuthal directions, respectively. The spatial coordinates in the corresponding directions are $x$, $r$ and $\theta$; the origin is centred on the upstream wall (figure~\ref{fig:sketch_domain}) so that the cavity axis is $r=0$. Finally, we shall define the hydrodynamics and acoustics problems using the same dimensionless parameters and variables as in our previous work \citep{Vincent2024scaling}.

The cavity, of diameter $D$ and length $L$, is entirely filled with a Newtonian fluid. The cavity is fitted with a plane circular transducer of unit diameter at $x=0$. The transducer, whose axis is aligned with $\boldsymbol{e_x}$, radiates an axisymmetric beam of linear acoustic pressure waves. As these waves travel towards the opposite cavity end, their amplitude decays for two reasons: diffraction, which also causes the beam radius to increase at a rate $S$ (figure~\ref{fig:sketch_domain}), and sound attenuation. The latter gives rise to an axisymmetric body force driving a streaming flow of velocity $\boldsymbol{u}$. The exponential decay rate of the acoustic pressure wave amplitudes is defined by the sound attenuation coefficient $N/L$, where $N$ is the ratio between the cavity length and the attenuation length of the acoustic pressure wave amplitude. Finally, all cavity walls absorb incident sound waves thus preventing the development of standing sound waves.

Whilst the sound field is characterised by small length scales and fast time scales, the streaming flow they give rise to is significantly slower than the speed of sound and involves length scales that greatly exceed the acoustic wavelength \citep{Lighthill1978,Orosco2022}. These scale differences allow to model streaming flows using the continuity and Navier-Stokes equations for an incompressible Newtonian fluid together with a body force \citep{Moudjed2014}:
\begin{equation} \label{eq:continuity_baseFlow}
	\bnabla \bcdot \boldsymbol{u} = 0 \, ,
\end{equation}
\begin{equation} \label{eq:NS_baseFlow}
	\frac{ \partial \boldsymbol{u} }{\partial t} + \left( \boldsymbol{u} \bcdot \bnabla \right) \boldsymbol{u} = - \bnabla p + \bnabla^2 \boldsymbol{u} + \boldsymbol{F_{ac}} \, ,
\end{equation}
in which $p$ refers to pressure, and
\begin{equation} \label{eq:acoustic_force}
\boldsymbol{F_{ac}} = Gr_{ac} \boldsymbol{ \widetilde{I} }
\end{equation}
is the body force coupling the acoustics and hydrodynamics problems. The spatial structure of $\boldsymbol{F_{ac}}$ is set by the normalised acoustic intensity $\boldsymbol{\widetilde{I}}$, which is bounded by unity for an unattenuated sound field. The magnitude of the forcing is defined by the Grashof number $Gr_{ac}$, which compares the magnitude of the acoustic force to the magnitude of the viscous forces at the scale of the transducer~\citep{ElGhani2021}. Finally, the differences between the acoustic and flow scales prevent any alteration of the forcing field due to the flow; therefore, $\boldsymbol{F_{ac}}$ is time-independent.

The cavity walls are impermeable boundaries and thus impose:
\begin{equation*}
\boldsymbol{u} = \boldsymbol{0}
\end{equation*}
at $x \in \left\{  0, L \right\}$ and $r = D/2$. Besides, the flow is axisymmetric and swirl-free, so that for all $x$ and $r$:
\begin{equation*}
u_{\theta} = \frac{\partial u_x}{\partial \theta} = \frac{\partial u_r}{\partial \theta} = \frac{\partial p}{\partial \theta} = 0\, ,
\end{equation*}
where $u_x$, $u_r$ and $u_{\theta}$ refer to the axial, radial and azimuthal velocity components, respectively. Axisymmetry of both $\boldsymbol{u}$ and $p$ further imposes
\begin{equation*}
u_r = \frac{\partial u_x}{\partial r} = \frac{\partial p}{\partial r} = 0
\end{equation*}
on the axis ($r=0$).

The transducer is modelled as a circular and plane baffled piston vibrating at a uniform velocity. The vibrating surface of the transducer is represented by an infinite number of acoustic point sources, each contributing to the acoustic field at any point $\left( x, r \right)$ of the fluid domain. The classical equations for a plane baffled piston in a semi-infinite domain, such as those found in \citet{Blackstock2000}, were adapted by \citet{Vincent2024scaling} to account for viscous attenuation. The authors showed that $\boldsymbol{ \widetilde{I} }$ can be obtained from a set of integrals defining the acoustic pressure $\widetilde{p}_{ac}$ and velocity $\boldsymbol{ \widetilde{u}_{ac} }$ at any point $\left( x, r \right)$ in the fluid domain:
\begin{subequations} \label{eq:Rayleigh_integral_all}
\begin{align}
\boldsymbol{ \widetilde{I} } &= \Re \left\{ \widetilde{p}_{ac} \, \left( \boldsymbol{ \widetilde{u}_{ac} }^* \right) \right\} \, , \label{eq:normalised_intensity}\\
\widetilde{p}_{ac} &= \mathrm{i} \frac{ k^2 S }{8 \times 1.22 \pi^2} \int_{0}^{1/2} \int_{0}^{2 \pi} \frac{ e^{- \mathrm{i} k \Vert \boldsymbol{d} \Vert } }{ \Vert \boldsymbol{d} \Vert } r' \mathrm{d}r' \mathrm{d}\theta' \, , \label{eq:Rayleigh_integral_pressure}\\
\boldsymbol{ \widetilde{u}_{ac} } &= \frac{1}{4 \pi} \int_{0}^{1/2} \int_{0}^{2\pi} \left( 1 + \mathrm{i} k \Vert \boldsymbol{d} \Vert \right) \frac{ e^{- \mathrm{i} k \Vert \boldsymbol{d} \Vert } }{ \Vert \boldsymbol{d} \Vert^2} \frac{ \boldsymbol{d} }{ \Vert \boldsymbol{d} \Vert }  r' \mathrm{d}r' \mathrm{d}\theta' \, , \label{eq:Rayleigh_integral_velocity}\\
k &= \frac{2.44 \pi}{S} - \mathrm{i} \frac{N}{L} \, , \label{eq:wave_number}
\end{align}
\end{subequations}
where $^{*}$ refers to the complex conjugate and $\Re$ to the real part. In equations~\eqref{eq:Rayleigh_integral_all}, ${\boldsymbol{d} = x \, \boldsymbol{e_x} + \left[ r - r' \cos \left( \theta' \right) \right] \boldsymbol{e_r} - r' \sin \left( \theta' \right) \boldsymbol{e_{\theta}}}$ is the distance between any point source on the transducer surface (located by the set of radial and azimuthal coordinates $r'$ and $\theta'$), and any point at $\left( x, r \right)$. Finally, $\boldsymbol{\widetilde{I}}$ is mainly shaped by the complex wave number $k$: the real part of $k$ sets the beam diffraction, whilst the imaginary part of $k$ involves the sound attenuation coefficient $N/L$ and thus adds attenuation to $\boldsymbol{\widetilde{I}}$.

Equations~\eqref{eq:Rayleigh_integral_all} define a beam-shaped axisymmetric acoustic field. As (i) the beam expands radially at a rate $S$ along $x$, and (ii) the beam radius matches the transducer radius (0.5 in dimensionless coordinates) at $x=0$, an approximate description of the beam radius $R_{beam}$ is:
\begin{equation} \label{eq:beam_radius}
    R_{beam} = 0.5 + S x \, ,
\end{equation}
from which we define the confinement ratio $C_R$:
\begin{equation} \label{eq:confinement_ratio}
C_R = \frac{ D/2 }{\left. R_{beam} \right|_{x=L} } \, .
\end{equation}
From now, we shall use $C_R$ as the main indicator of the expected radial confinement of the flow; small (respectively large) $C_R$ values correspond to cases of high (respective weak) confinement.

To investigate the effect of confinement on the flow stability, we shall consider six different cavity sizes by varying either $C_R$ or $N$ (table~\ref{tab:parameter_values}). We take $C_R \in \left[ 1, 6 \right]$ to be consistent with the experiments of \citet{Kamakura1996}, \citet{Mitome1998} and \citet{Moudjed2014} for which $C_R=2.9$, $8.5$ and $4.1$, respectively.  
\begin{table}
    \begin{center}
        \begin{tabular}{ccccc}
        $C_R$ & $N$ & $S$ & $L$ & $D$ \\
        1 & 0.25 & 0.03 & ~83.3 & ~~6.02\\
        2 & 0.25 & 0.03 & ~83.3 & ~12.04\\
        3 & 0.25 & 0.03 & ~83.3 & ~18.06\\
        4 & 0.25 & 0.03 & ~83.3 & ~24.08\\
        6 & 0.25 & 0.03 & ~83.3 & ~36.12\\
        6 & 1 & 0.03 & 333.3 & 126.51\\
        \end{tabular}
    \end{center}
\caption{Values of the parameters defining each setup. The parameter $C_R$ is defined as the ratio between the cavity radius $D/2$ and the approximate beam radius evaluated at $x=L$ using equation~\eqref{eq:beam_radius}. The values of $S$ and of the acoustic pressure attenuation coefficient $N/L$ are chosen to match the 2 MHz water experiments of \citet{Moudjed2014}.}
\label{tab:parameter_values}
\end{table}
As these experiments were made for $N < 1$, we set $N=0.25$ for five cases. Nevertheless, from a process standpoint, it is preferable to have $N \geq 1$ to ensure that most of the acoustic kinetic energy radiated by the transducer is used to create flow kinetic energy \citep{Vincent2024scaling}. We shall thus investigate a last case for which $\left( N, C_R \right) = \left( 1, 6 \right)$ (table~\ref{tab:parameter_values}). For all six cases, the parameters $S = 0.03$ and $N/L = 0.003$ are kept constant; these values are based on the 2~MHz water experiments of \citet{Moudjed2014} and \citet{Moudjed2015}. Therefore, the structure of $\boldsymbol{F_{ac}}$ in our work is consistent with the forcing fields of former experiments. These experiments covered a wide range of $Gr_{ac}$; it typically ranged from ${5.6 \times 10^2}$ \citep{Mitome1998} to ${1.2 \times 10^5}$ \citep{Moudjed2014}. In the present work, we shall adjust $Gr_{ac}$ on a case-by-case basis to identify the instability onset.

\subsection{Linear stability analysis}
\label{subsec:linear_stability_analysis}

Although the base flow is 2D-2C (two-dimensional, two-component), it may nevertheless be destabilised by non-axisymmetric 3D-3C perturbations under certain forcing and geometrical conditions. We shall therefore decompose the flow fields into a steady axisymmetric part corresponding to the base flow, and an infinitesimal time-dependent perturbation:
\begin{equation} \label{eq:velocity_decomposition}
\boldsymbol{u} = \boldsymbol{U}\left( x, r \right) + \boldsymbol{u'}\left( x, r, \theta, t \right) \, ,
\end{equation}

\begin{equation} \label{eq:pressure_decomposition}
p = P\left( x, r \right) + p'\left( x, r, \theta, t \right) \, .
\end{equation}
The perturbations $\left( \boldsymbol{u'}, p' \right)$ are purely hydrodynamic i.e., they dot not correspond to any acoustic signals as feedback effects of the flow on the acoustic fields are discarded. The linear stability equations are then obtained by inserting the decompositions \eqref{eq:velocity_decomposition} and \eqref{eq:pressure_decomposition} into the Navier-Stokes~\eqref{eq:NS_baseFlow} and continuity~\eqref{eq:continuity_baseFlow} equations, and by linearising the resulting equations around the base flow. Since $\left( \boldsymbol{U}, P \right)$ satisfy equations~\eqref{eq:NS_baseFlow} and~\eqref{eq:continuity_baseFlow}, the linear equations for $\left( \boldsymbol{u'}, p' \right)$ are thus:
\begin{equation} \label{eq:linearised_NS}
\frac{ \partial \boldsymbol{u'} }{ \partial t } + \left( \boldsymbol{U} \bcdot \bnabla \right) \boldsymbol{u'} + \left( \boldsymbol{u'} \cdot \bnabla \right) \boldsymbol{U} = - \bnabla p' + \bnabla^2 \boldsymbol{u'} \, ,
\end{equation}

\begin{equation} \label{eq:linearised_continuity}
\bnabla \bcdot \boldsymbol{u'} = 0 \, .
\end{equation}
As the base flow is invariant along $\boldsymbol{e_{\theta}}$, a generic three-dimensional perturbation $\boldsymbol{q'}$ is expressed using Fourier series:
\begin{equation} \label{eq:fourier_decomposition}
\boldsymbol{q'} \left( x, r, \theta, t \right) = \sum_{m=-\infty}^{+\infty} \widehat{ \boldsymbol{q'} }_m \left( x, r, t \right)e^{\textrm{i} m \theta} \, ,
\end{equation}
where the integer $m$ is the azimuthal wavenumber. Since equations~\eqref{eq:linearised_NS} and \eqref{eq:linearised_continuity} are linear, the LSA is thus decoupled for each $m$. The linearity of equations ~\eqref{eq:linearised_NS} and \eqref{eq:linearised_continuity} decouples the Fourier modes, so that the LSA (Linear Stability Analysis) equations can be solved individualy for each $m$. Finally, the perturbations satisfy the same boundaries conditions as the base flow, except at $r=0$ where $m$-dependent kinematic conditions are imposed to allow for three-dimensional perturbations to develop whilst avoiding the geometrical singularity \citep{Blackburn2004}.

The linear stability problem is addressed using a timestepper approach. From equations~\eqref{eq:linearised_NS} and~\eqref{eq:linearised_continuity}, the perturbation at a time $t + \tau$ results from the action of an evolution operator $\mathcal{A}(\tau)$ applied to the same perturbation at a previous time $t$~\citep{Barkley2008}:
\begin{equation*}
\boldsymbol{q'}\left( t + \tau \right) = \mathcal{A}\left( \tau \right) \boldsymbol{q'}\left( t \right) \, .
\end{equation*}
The complex eigenvalue $\zeta_k$ of $\mathcal{A}\left( \tau \right)$, associated with the eigenvector $\boldsymbol{q'}_k$, is linked to the eigenvalue $\left( \sigma_k + \mathrm{i} \omega_k \right)$ of the linear problem \eqref{eq:linearised_NS}-\eqref{eq:linearised_continuity} through an exponential mapping:
\begin{equation*}
\zeta_k = e^{\left( \sigma_k + \mathrm{i} \omega_k  \right) \tau} \, ,
\end{equation*}
from which the growth rate $\sigma_k$ and frequency $\omega_k$ can be retrieved:
\refstepcounter{equation}
$$
  \sigma_k = \frac{ \textrm{ln}\left( \vert \zeta_k \vert \right) }{\tau}, \quad
  \omega_k = \frac{\varphi_k}{\tau} \, ,
  \eqno{(\theequation{\mathit{a},\mathit{b}})}
$$
where $\varphi_k$ is the complex argument of $\zeta_k$. A given base flow $\left( \boldsymbol{U}, P \right)$ is said to be unstable if, for any $m$, there is at least one perturbation $\left( \boldsymbol{u'}, p' \right)$ for which $\sigma > 0$ (or, equivalently, $\vert \zeta \vert > 1$). Conversely, the base flow is stable if the growth rates associated with all the eigenvalues of the
linear problem are strictly negative. The minimum value of $Gr_{ac}$ for which $\sigma = 0$ is reached is called the critical acoustic Grashof number $Gr_{ac}^c$ and indicates the onset of an instability.

\section{Computational methodology}
\label{sec:computational_methodology}

\subsection{Numerical methods}
\label{subsec:numerical_methods}

Our work involves four different numerical calculations. First, we calculate $\boldsymbol{F_{ac}}$ at the grid points in the fluid domain. Second, we compute the steady axisymmetric 2D-2C streaming flow driven by $\boldsymbol{F_{ac}}$ by time-stepping equations~\eqref{eq:NS_baseFlow}-\eqref{eq:continuity_baseFlow} until a steady-state is reached. Third, we perform the LSA of the axisymmetric 2D-2C flow for each case listed in table~\ref{tab:parameter_values}. The leading eigenmode and its eigenvalue are computed by solving the eigenvalue problem for $0 \leq m \leq 10$. This allows to determine $Gr_{ac}^c$ and identify the shape of the unstable perturbation for each case. Fourth, nonlinear 3D-3C unsteady simulations are run at slightly supercritical $Gr_{ac}$ for two purposes: (i) to validate the LSA, and (ii) to determine the bifurcation nature (sub- or supercritical). The latter is inferred by modelling the growth and saturation of a perturbation near $Gr_{ac}^c$ with a Stuart-Landau equation~\citep{Landau1987}, the parameters of which being evaluated from time series of the flow variables~\citep{Sheard2004}.

The forcing field is obtained by numerically evaluating the integrals in equation~\eqref{eq:Rayleigh_integral_all} at the grid points. The transducer is discretised by a homogeneous distribution of $N_s$ point sources in both the radial and the azimuthal directions. The total number of point sources used to estimate $\boldsymbol{ \widetilde{I} }$ (hence $\boldsymbol{F_{ac}}$) at any grid point in the fluid domain is thus $N_s^2$.

Both the steady axisymmetric 2D-2C and the unsteady 3D-3C flows are computed using the open-source spectral-element code Semtex \citep{Blackburn2019}. The $\left(x, r \right)$ plane is split into non-overlapping elements, inside which the flow variables are expanded using a tensor product of Lagrange polynomials defined on the Gauss-Lobatto-Legendre points~\citep{Blackburn2004}. Convergence of the solution is then achieved by increasing the polynomial degree $N_p$ of the expansion basis. For the 3D-3C unsteady simulations, the azimuthal direction is discretised using a Fourier spectral method~\citep{Blackburn2004}.

\begin{figure}
    \centering
    \includegraphics[scale=0.5]{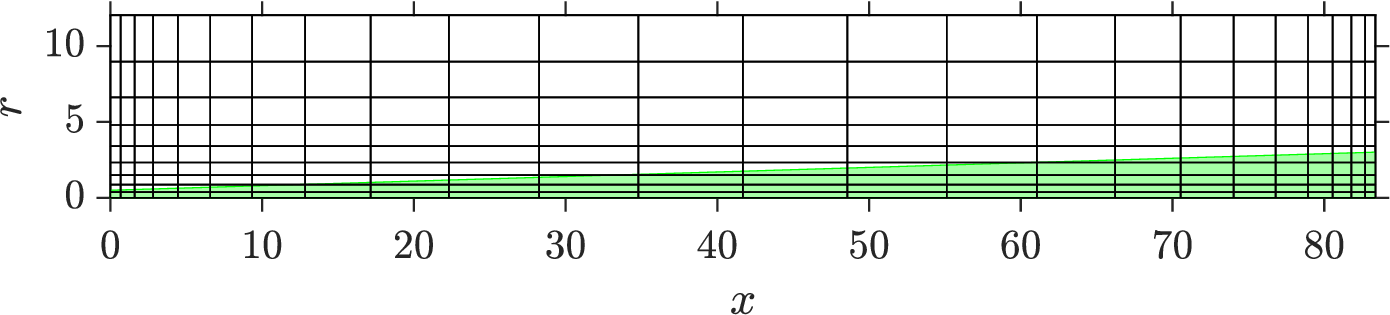}
    \caption{Typical mesh used to discretise the fluid domain. The circular transducer is placed at $x=0$ and its axis is aligned with the $r=0$ line. The beam, of approximate radius $R_{beam}$ (equation~\eqref{eq:beam_radius}), is shown in green. The displayed mesh corresponds to the $\left(N, C_R\right) = \left( 0.25, 4 \right)$ case.}
    \label{fig:mesh_case3}
\end{figure}

The $\left(x, r\right)$ plane is discretised using a grid of quadrilateral elements (figure~\ref{fig:mesh_case3}).
For instance, the mesh used for the $\left( N, C_R \right)=\left( 0.25, 4 \right)$ comprises 24 and 9 elements in the axial and radial directions, respectively. For each case, the grid is refined near the upstream and downstream walls, so that the ratio of the lengths of two neighbouring elements does not exceed 1.35. In the radial direction, the element height is increased by a factor 1.3 as the distance from the axis is increased. The grid density in the radial direction is set such that there is at least one element in $0 \leq r \leq 0.5$. We also adapted the total number of elements to keep a similar spatial resolution of $\boldsymbol{F_{ac}}$ in the very-near acoustic field ($x \leq 1$) for each case (table~\ref{tab:mesh_characteristics}).

\begin{table}
    \begin{center}
        \begin{tabular}{ccccc}
        $N$ & $C_R$ & \thead{Elements in \\axial direction} & \thead{Elements in\\ radial direction} & \thead{Total number \\of elements} \\
        0.25 & 1 & 24 & 5 & 120\\
        0.25 & 2 & 24 & 7 & 168\\
        0.25 & 3 & 24 & 9 & 216\\
        0.25 & 4 & 24 & 9 & 216\\
        0.25 & 6 & 24 & 10 & 240\\
        1 & 6 & 34 & 14 & 476\\
        \end{tabular}
    \end{center}
\caption{Characteristics of the meshes used for each case. See table~\ref{tab:parameter_values} for a complete list of the computational parameters defining each case.}
\label{tab:mesh_characteristics}
\end{table}

Equations~\eqref{eq:continuity_baseFlow}-\eqref{eq:NS_baseFlow} are discretised in time using the operator splitting scheme of \citet{Karniadakis1991}. The scheme integrates explicitly the inertia terms of equation~\eqref{eq:NS_baseFlow}, and implicitly its viscous terms. For the unsteady 3D-3C simulations, the temporal integration scheme is used in its third-order formulation. For the 2D-2C base flow calculations, as only steady solutions of equations~\eqref{eq:continuity_baseFlow}-\eqref{eq:NS_baseFlow} are sought, the scheme is used in its first-order version. A steady state is then considered to be reached when the time series of the flow variables, recorded at different locations within the fluid domain, do not display variations with time up to a precision of seven significant figures.

We carry out the LSA using the open-source code DOG (Direct Optimal Growth, \citet{Barkley2008}), which is based on a time-stepper method and a spectral element spatial discretisation. The code marches forward in time the linear stability equations~\eqref{eq:linearised_NS}-\eqref{eq:linearised_continuity} using the second-order formulation of the operator splitting scheme of \citet{Karniadakis1991}. The eigenvalues and associated eigenvectors of the stability problem are then recovered by applying the iterative algorithm of \citet{Barkley2008}.

Finally, for a given set of values of $N$, $C_R$ and $Gr_{ac}$, the initial condition for the 2D-2C flow calculations is either set to $\boldsymbol{u}=0$, or to a previous solution obtained for the same $\left( N, C_R \right)$ but for a different $Gr_{ac}$ to accelerate the convergence towards a steady state. The unsteady 3D-3C simulations at slightly supercritical $Gr_{ac}$ are initialised with the steady axisymmetric base flow duplicated along $\boldsymbol{e}_{\theta}$, plus a Gaussian white noise. The noise is evenly distributed over all the azimuthal wavenumbers, thus ensuring that the initial disturbance is not biased towards a preferred shape. The standard deviation of the white noise was set to 1 for the 3D-3C unsteady simulations run for $C_R = 6$. As we reduced $C_R$, such standard deviation yielded very large initial disturbances, making the early exponential growth of the perturbation impossible to observe. We thus reduced the noise standard deviation to 0.1 for $C_R \in \left\{ 3, 4  \right\}$, and to 0.01 for $C_R \in \left\{ 1, 2 \right\}$.

\subsection{Grid sensitivity analysis}
\label{sec:grid_sensitivity_analysis}

\begin{figure}
    \centering
    \includegraphics[scale=0.5]{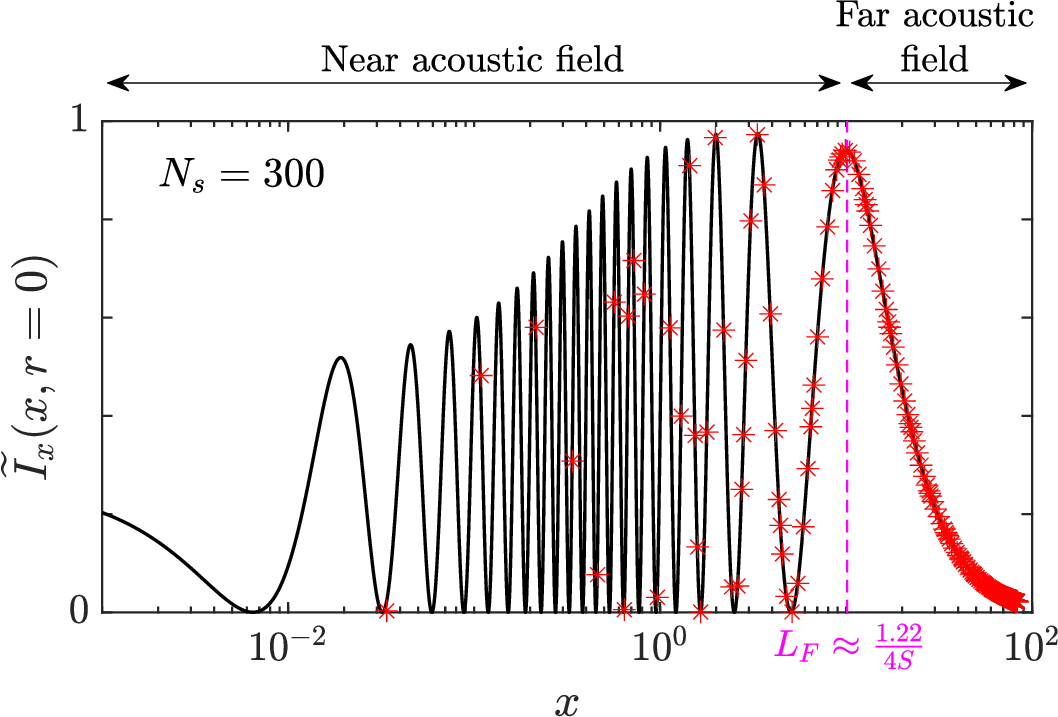}
    \caption{On-axis normalised acoustic intensity profile $\widetilde{I}(x,r=0)$ for $\left( N, C_R \right)=\left( 0.25, 4 \right)$. The analytical profile, proposed by \citet{Vincent2024scaling}, is shown in black. The transition between the near and far acoustic fields occurs at the last intensity peak located at approximately the Fresnel distance $L_F \approx 1.22 / \left( 4 S \right)$ (purple dashed vertical line). The values obtained by numerically evaluating equations~\eqref{eq:Rayleigh_integral_all} at the collocation points of the mesh (with an expansion basis of degree $N_p=8$ in each element) are represented by red stars. The numerical points are obtained by discretising the transducer with $N_s = 300$ point sources in both the radial and the azimuthal directions.}
    \label{fig:intensity_profile}
\end{figure}

We assessed the suitability of the grids discretising the transducer and the fluid domain. For the transducer discretisation, we previously derived an analytical expression for the on-axis intensity $\widetilde{I}_x(x,r=0)$, which we compared to the numerical evaluation of equation~\eqref{eq:Rayleigh_integral_all} using the methodology outlined in \S~\ref{subsec:numerical_methods} \citep{Vincent2024scaling}. We showed that discretising the transducer with $N_s=300$ grid points in both the radial and azimuthal directions ensures a local error of less than 1~\% between the analytical and numerical intensity profiles on the axis (represented by the black curve and red stars, respectively, in figure~\ref{fig:intensity_profile}). All the results presented in this work were thus obtained with $N_s=300$.

As shown in figure~\ref{fig:intensity_profile}, the forcing field features strong gradients in the near acoustic field. These sharp intensity variations occur over very short length scales and would require an extremely fine grid to be accurately captured. As these length scales are significantly smaller than those of the studied streaming flow \citep{Vincent2024scaling}, the very near acoustic field (i.e., $x \leq 1$) is undersampled to keep the flow calculations tractable. We nevertheless performed $p$-refinement to determine the minimum $N_p$ above which the variations of the leading eigenvalue remained below 0.5~\%.
\begin{table}
    \begin{center}
        \begin{tabular}{ccc}
        $N_p$ & $\sigma_{max}$ & Relative difference (\%)\\
        \\
        5 & 0.10249 & 33.11 \\
        6 & 0.07379 & ~~4.17 \\
        7 & 0.07792 & ~~1.20 \\
        8 & 0.07695 & ~~0.05\\
        9 & 0.07710 & ~~0.13\\
        10 & 0.07709 & ~~0.12\\
        11 & 0.07696 & ~~0.05\\
        12 & 0.07702 & ~~0.03\\
        13 & 0.07701 & ~~0.02\\
        14 & 0.07700 & $-$\\
        \end{tabular}
    \end{center}
\caption{Evolution of the leading mode growth rate $\sigma_{max}$ with the polynomial degree $N_p$ of the expansion basis. The growth rates are obtained for $\left( N, C_R \right)=\left( 0.25, 4 \right)$ with $Gr_{ac} = 7000$ and $m=2$. The error is relative to the value of $\sigma_{max}$ obtained for $N_p = 14$.}
\label{tab:mesh_convergence}
\end{table}
An example of grid sensitivity analysis for the ${\left( N, C_R \right) = \left( 0.25, 4 \right)}$ case at $Gr_{ac} = 7000$ (the largest $Gr_{ac}$ for that setup) is shown in table~\ref{tab:mesh_convergence}. The leading perturbation is an unstable non-oscillatory $m=2$ mode. We chose $N_p=8$, since variations of $\sigma$ with respect to the most refined grid were at most $\approx 0.1$~\% for larger $N_p$. We then repeated this $p$-refinement study for all $\left( N, C_R \right)$ to ensure that our stability results are mesh-independent.

For the 3D-3C unsteady simulations, $\boldsymbol{u}$ and $p$ were expanded along $\boldsymbol{e_{\theta}}$ using $N_F$ Fourier modes. We assessed the effect of $N_F$ by computing the flow kinetic energy over the entire cavity after the perturbation had saturated. For $\left( N, C_R \right) = \left( 0.25, 1 \right)$ with $Gr_{ac} = 15500$ for instance, the relative difference between the kinetic energies computed with either $N_F = 32$ or 64 was $\approx 0.03$~\%. Even smaller differences were obtained for the $C_R \in \left\{ 4, 6 \right\}$ cases at supercritical forcings. The 3D-3C unsteady simulations were thus performed with $N_F = 32$ for all cases.

Finally, we adapted the time step $\Delta t$ to each case so that the Courant-Friedrichs-Lewy (CFL) number remained below unity to ensure the numerical stability. For instance, the $\left( N, C_R \right) = \left( 0.25, 4 \right)$ case at $Gr_{ac} = 7000$ and with $N_p = 8$ was time-stepped with $\Delta t = 10^{-4}$, yielding a maximum CFL number of 0.476.

\section{Two-dimensional axisymmetric base flow}
\label{sec:base_flow}

\subsection{Impinging jet structure of the streaming flow in a weakly confined setting}

\begin{figure}
    \centering
    \includegraphics[width=0.9\textwidth]{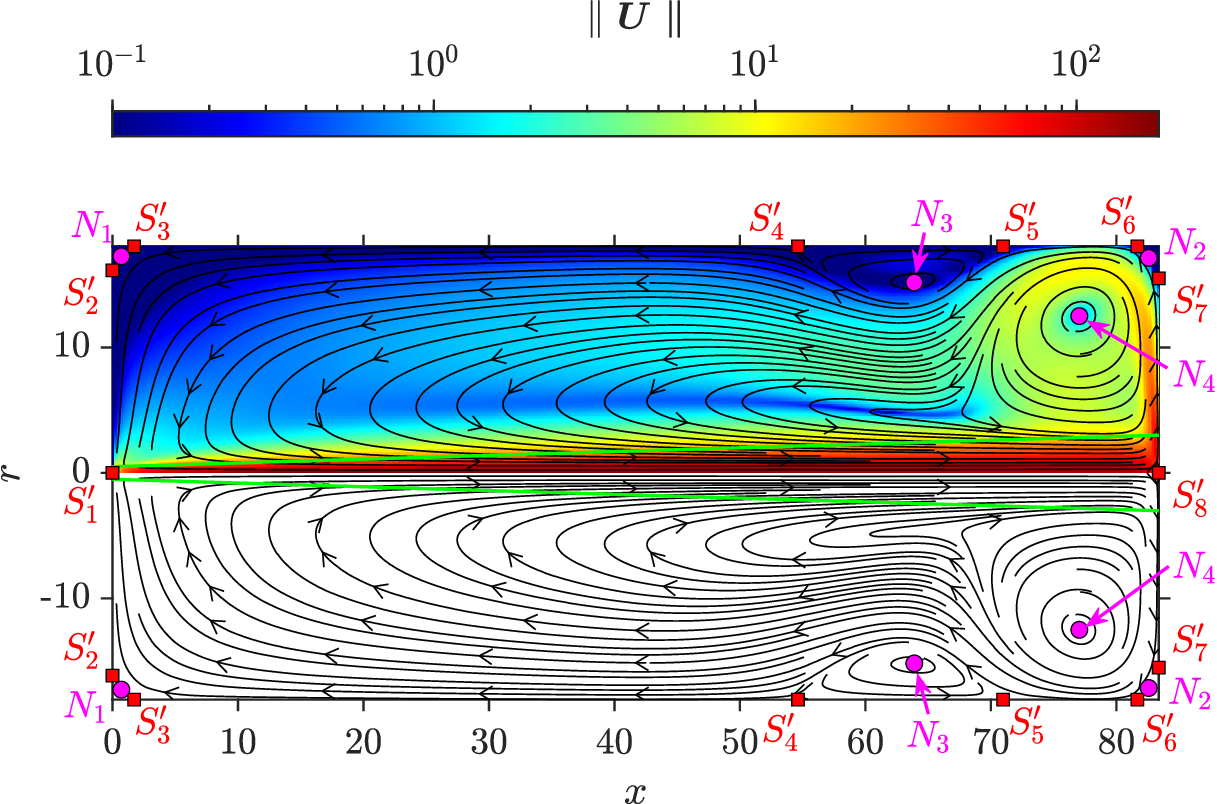}
    \caption{Map of the steady velocity magnitude $\Vert \boldsymbol{U} \Vert$ computed for ${\left( N, C_R \right) = \left( 0.25, 6 \right)}$ at $Gr_{ac} = 6.4\times 10^3$. The streamlines are displayed together with the base flow critical points: half saddles (zero wall skin friction, red squares) are labelled as $S'_i$ and $N_i$ are nodes ($\boldsymbol{U} = \boldsymbol{0}$, purple discs). The green line depicts the approximate beam radius given by equation~\eqref{eq:beam_radius}. The base flow is axisymmetric and is reflected about the $r=0$ axis for the sake of visualisation.}
    \label{fig:base_flow_case_2}
\end{figure}

Before studying its stability, we shall first discuss the steady two-dimensional and axisymmetric base flow. Figure~\ref{fig:base_flow_case_2} shows a typical base flow considered for LSA, with the ${\left( N,  C_R \right)=\left( 0.25, 6 \right)}$ case taken as illustrative example.
The main flow features are identified by the patterns of streamlines together with critical point theory~\citep{Chong1990}. The critical point analysis is commonly used to identify structures within complex 3D flows \citep{Hunt1978}. It was first introduced to analyse the wall-stress fields of flows around bluff bodies in wind-tunnels but we showed that its application to the complex flow field produced by 3D DNS (Direct Numerical Simulations) of separated flows around bluff bodies \citep{Dousset2010,Dousset2012} or complex conduits \citep{Potherat2018} offered a powerful way to reliably extract the main structures governing the flow's dynamics. The main idea is to extract 2D streamlines or stresslines along symmetry planes or solid walls, and to classify the critical points, where both components of velocity or stress cancel, into nodes and saddles. First, the theoretical constraint on the number of saddles and nodes offers a convenient way to validate the numerical simulations \citep{Hunt1978}. Second, these points offers a simple parametrisation of the entire flow field. When a flow transitions between regimes, topological changes are straightforwardly captured by changes in the number of critical points, and the nature of these points often points to the physical mechanism underpinning the transition. We use it here for the first time in the context of stability analysis, and apply it separately to the base flow and to the perturbations.

The critical points where $\boldsymbol{U} = \boldsymbol{0}$ are shown in purple (labelled $S_i$ for saddles and $N_i$ for nodes, respectively) and points of zero skin friction on the solid boundaries, displayed in red, are half saddles labelled as $S'_i$. The base flow field typically comprises a jet aligned with the cavity axis. Being driven by the acoustic forcing, the jet originates at $S'_1$ and impinges the wall facing the transducer. The jet impingement at $S'_8$ then gives rise to a large roll centred on $N_4$. This roll features weak velocity amplitudes, and spreads over the entire space available between the jet and the lateral cavity wall. The presence of this roll causes the return flow to separate from and reattach to the wall at $S'_5$ and $S'_4$, respectively. A recirculation bubble centred at $N_3$ lies between these two points.

The distribution and type of critical points within a flow field obey a strict topological rule. More precisely, the number of nodes $\Sigma_{N}$, of half-nodes $\Sigma_{N'}$, of saddles $\Sigma_{S}$ and half-saddles $\Sigma_{S'}$ in a slice of $\boldsymbol{U}$ are related to each other through~\citep{Hunt1978,Foss2004}:
\begin{equation}\label{eq:topological_rule}
    2\Sigma_{N} + \Sigma_{N'} - 2\Sigma_{S} - \Sigma_{S'} = \chi \, ,
\end{equation}
where $\chi$ is a parameter defining the topology of the surface in which the streamline patterns are studied ($\chi = 2$ for the plane shown in figure~\ref{fig:base_flow_case_2} \citep{Foss2004}). Inspection of the streamlines in figure~\ref{fig:base_flow_case_2} gives:
\begin{equation*}
    \Sigma_{N} = 8\, , \quad \Sigma_{N'} = 0 \, , \quad \Sigma_{S} = 0 \, , \quad \Sigma_{S'} = 14 \, ,
\end{equation*}
so that the left-hand side of equation~\eqref{eq:topological_rule} is indeed equal to $\chi = 2$. The topological rule~\eqref{eq:topological_rule} is thus satisfied, meaning either that the entire flow topology is captured, or at least that unresolved parts of the flow are consistently cut off. Indeed, only one out of the infinite series of corner vortices theorised by \citet{Moffatt1964} is captured in each corner. Increasing the resolution would reveal a greater number of them. By ``consistently", we mean that with each additional vortex, a node and two half-saddles would add to the tally of critical points, without invalidating equation (\ref{eq:topological_rule}).

\subsection{Effect of confinement on the base flow topology and on the jet velocity}

Reducing the cavity size by either varying $N$ or $C_R$ may affect the 2D-2C base flow structure.
\begin{figure}
    \centering
    \includegraphics[width=0.99\textwidth]{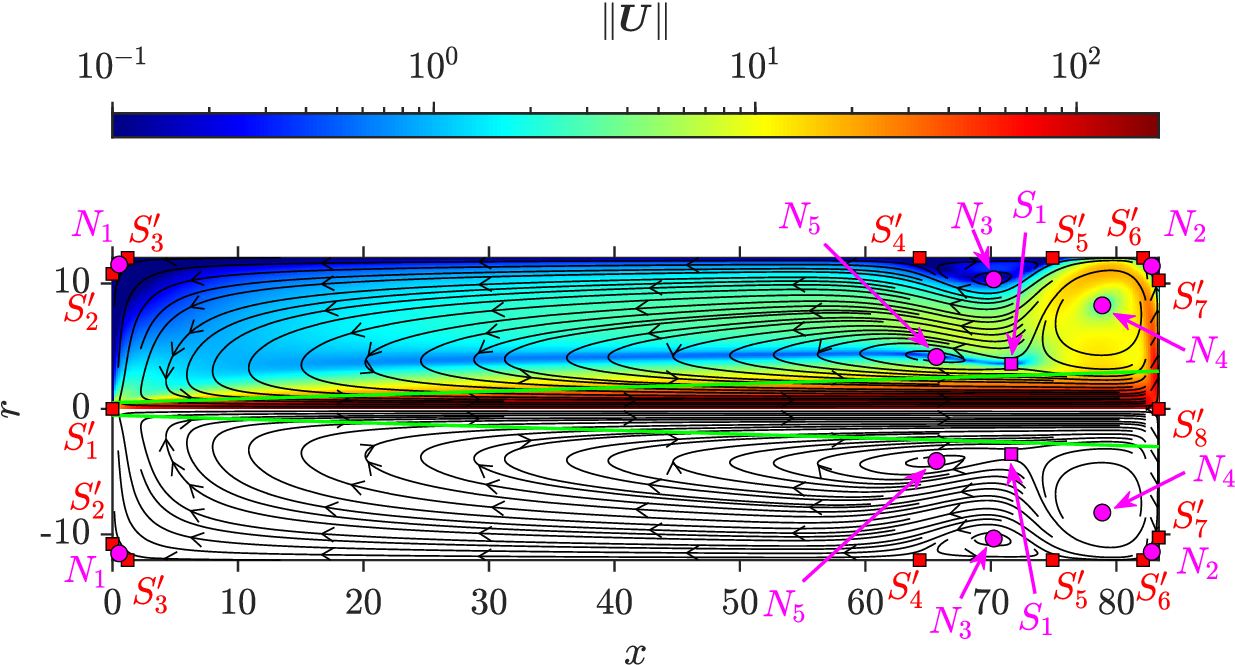}
    \caption{Map of the steady velocity magnitude $\Vert \boldsymbol{U} \Vert$ computed for ($N = 0.25$, $C_R=4$) for $Gr_{ac} = 6.4\times 10^3$. The streamlines are displayed together with the base flow critical points: half saddles (zero wall skin friction, red squares) are labelled as $S'_i$, $N_i$ are nodes ($\boldsymbol{U} = \boldsymbol{0}$, purple discs) and $S_i$ are saddles ($\boldsymbol{U} = \boldsymbol{0}$, purple squares). The green line depicts the approximate beam radius given by equation~\eqref{eq:beam_radius}. The base flow is axisymmetric and is reflected about the $r=0$ axis for the sake of visualisation.}
    \label{fig:baseFlow_case3_GrAc_6400}
\end{figure}
However, any topological modification of the base flow shall comply with the topological rule~\ref{eq:topological_rule}. Figure~\ref{fig:baseFlow_case3_GrAc_6400} shows the steady axisymmetric field obtained at the same $Gr_{ac}$ as in figure~\ref{fig:base_flow_case_2}, but for the smaller value of $C_R=4$. The increased radial confinement yields two additional critical points: a node $N_5$ at $x \approx 66$ and a saddle $S_1$ at $x \approx 72$. This new critical point combination keeps the left-hand side of the topological rule~\eqref{eq:topological_rule} identical to the $C_R=6$ case (figure~\ref{fig:base_flow_case_2}); the base flow for $C_R = 4$ shown in figure~\ref{fig:baseFlow_case3_GrAc_6400} is thus again valid from a topological perspective. Asides of these examples, a similar change of the base velocity field topology has only been observed when reducing $C_R$ from 2 to 1 and for forcing magnitudes close to their respective $Gr_{ac}^c$.

\begin{figure}
    \centering
    \includegraphics[width=0.9\textwidth]{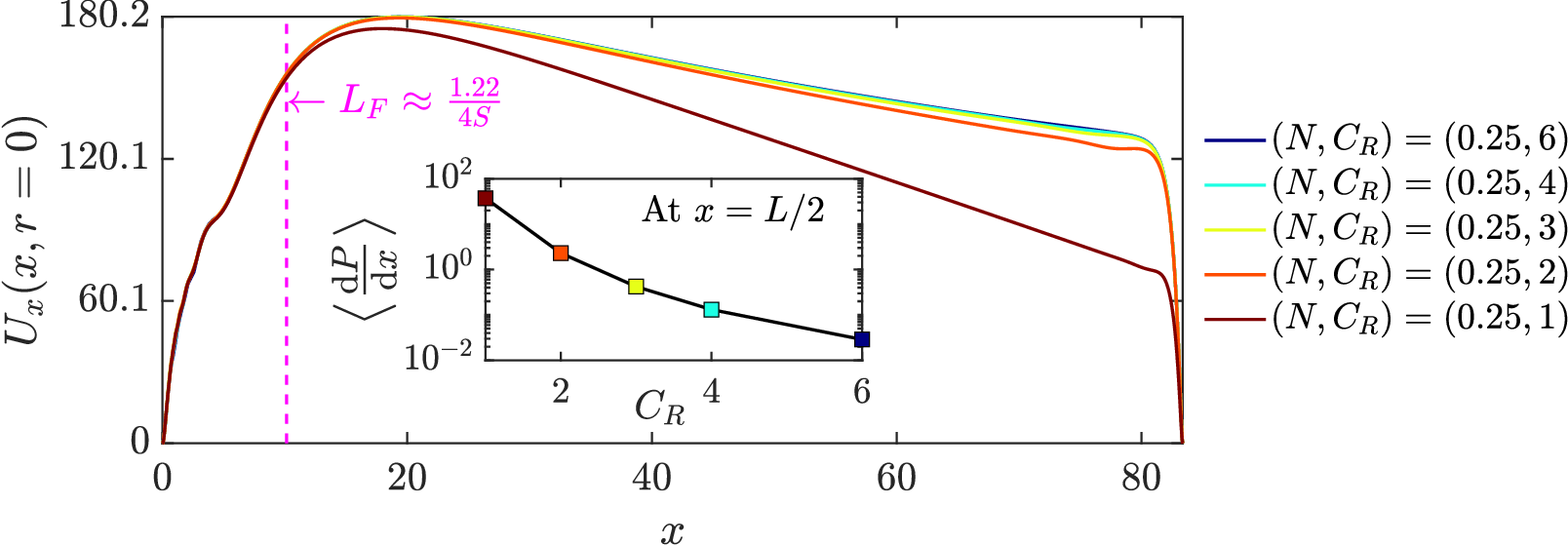}
    \caption{Profiles of the on-axis velocity $U_x(x,r=0)$ along the jet illustrating the effect of flow confinement on the jet velocity for $N=0.25$ and $Gr_{ac} = 6400$. The profiles are obtained for $N = 0.25$ and $1 \leq C_R \leq 6$. The dashed vertical line corresponds to the Fresnel distance $L_F$ marking the transition between the near ($x < L_F$) and the far ($x > L_F$) acoustic fields. Inset: variations with $C_R$ of the streamwise pressure gradient $\left\langle \mathrm{d}P / \mathrm{d} x \right\rangle$ of the base flow at the mid cavity length $L/2$ and averaged over the cross-sectional area.}
    \label{fig:longitudinalVelocityProfiles_cases2_to_6_GrAc6400}
\end{figure}

Besides potentially modifying the base flow topology, reducing $C_R$ also affects the jet velocity. Figure~\ref{fig:longitudinalVelocityProfiles_cases2_to_6_GrAc6400} shows several profiles of the centreline base velocity ${U_x \left(x,r=0 \right)}$ computed for $1 \leq C_R \leq 6$ and $Gr_{ac} = 6400$. In a weakly-confined setting, the jet strongly accelerates over $0 \leq x \lesssim 20$ for all $C_R$ as a result of a balance between $\boldsymbol{F_{ac}}$ and the inertia forces~\citep{Moudjed2014,Vincent2024scaling}. Decreasing $C_R$ enhances the jet deceleration at larger $x$. For $2 \leq C_R \leq 6$, this confinement-induced jet deceleration is marginal: ${U_x \left(x,r=0 \right)}$ differs by at most $5.5$~\%, and its decrease along the jet results from a balance between $\boldsymbol{F_{ac}}$ and viscous forces~\citep{Vincent2024scaling}. However, the jet deceleration is significantly stronger for $C_R = 1$: near the jet impingement, the centreline velocity is 50~\% less than for the $C_R \geq 2$ cases. This enhanced jet deceleration is caused by the build-up of a streamwise pressure gradient as $C_R$ is decreased (inset of figure~\ref{fig:longitudinalVelocityProfiles_cases2_to_6_GrAc6400}). This pressure force acts against the acoustic forcing and drives the return flow so that the net mass flux through a cross section is zero~\citep{Rudenko1998}.

\section{Stability analysis} \label{sec:stability_analysis}

\subsection{Non-oscillatory instabilities in weakly-confined settings}

\begin{figure}
    \centering
    \includegraphics[width=0.99\textwidth]{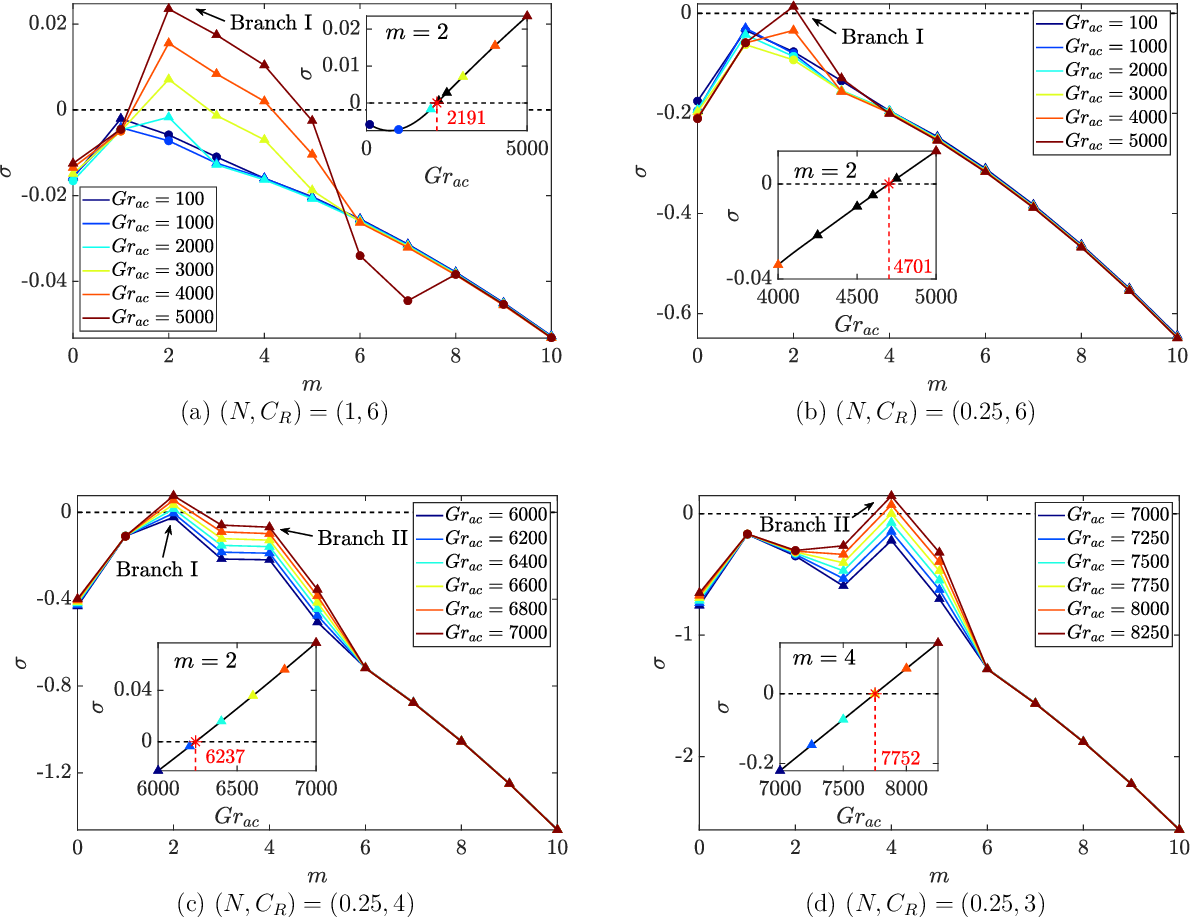}
    \caption{Growth rate $\sigma$ of the leading eigenmode as a function of the azimuthal wavenumber $m$ for $3 \leq C_R \leq 6$. Oscillatory and non-oscillatory modes are represented by filled circles and triangles, respectively. The insets show the evolution of $\sigma$ with $Gr_{ac}$ for the indicated value of $m$. The critical Grashof number $Gr_{ac}^c$, obtained through spline interpolation of $\sigma$, is reported in red. The filled black symbols in the inset of (b) are additional eigenvalue computations made to improve the interpolation accuracy near ${\sigma = 0}$.}
    \label{fig:sigma_vs_m_cases_1_to_4}
\end{figure}

We now turn to the linear stability analysis of the steady 2D-2C flows similar to those shown in figures~\ref{fig:base_flow_case_2} and~\ref{fig:baseFlow_case3_GrAc_6400}. The variations of the growth rate of the leading eigenmode $\sigma$ with $Gr_{ac}$ and $m$ for $N \in \left\{  0.25, 1 \right\}$ and $4 \leq C_R \leq 6$ are shown in figure~\ref{fig:sigma_vs_m_cases_1_to_4}.
For $4 \leq C_R \leq 6$, the same non-oscillatory $m=2$ mode is found to destabilise the flow; we shall denote the associated mode branch as ``Branch I". The unstable mode branch remains the same for ${4 \leq C_R \leq 6}$, but $Gr_{ac}^c$ increases from $2191$ for $\left( N, C_R \right) = \left( 1, 6 \right)$ to $6237$ for $\left( N, C_R \right) = \left( 0.25, 4 \right)$. Confining the streaming jet flow thus has a stabilising effect, similarly to confined flows past bluff bodies~\citep{Mondal2023}.

\begin{figure}
    \centering
    \includegraphics[trim={0cm, 2.5cm, 0cm, 1.5cm}, clip, width=0.99\textwidth]{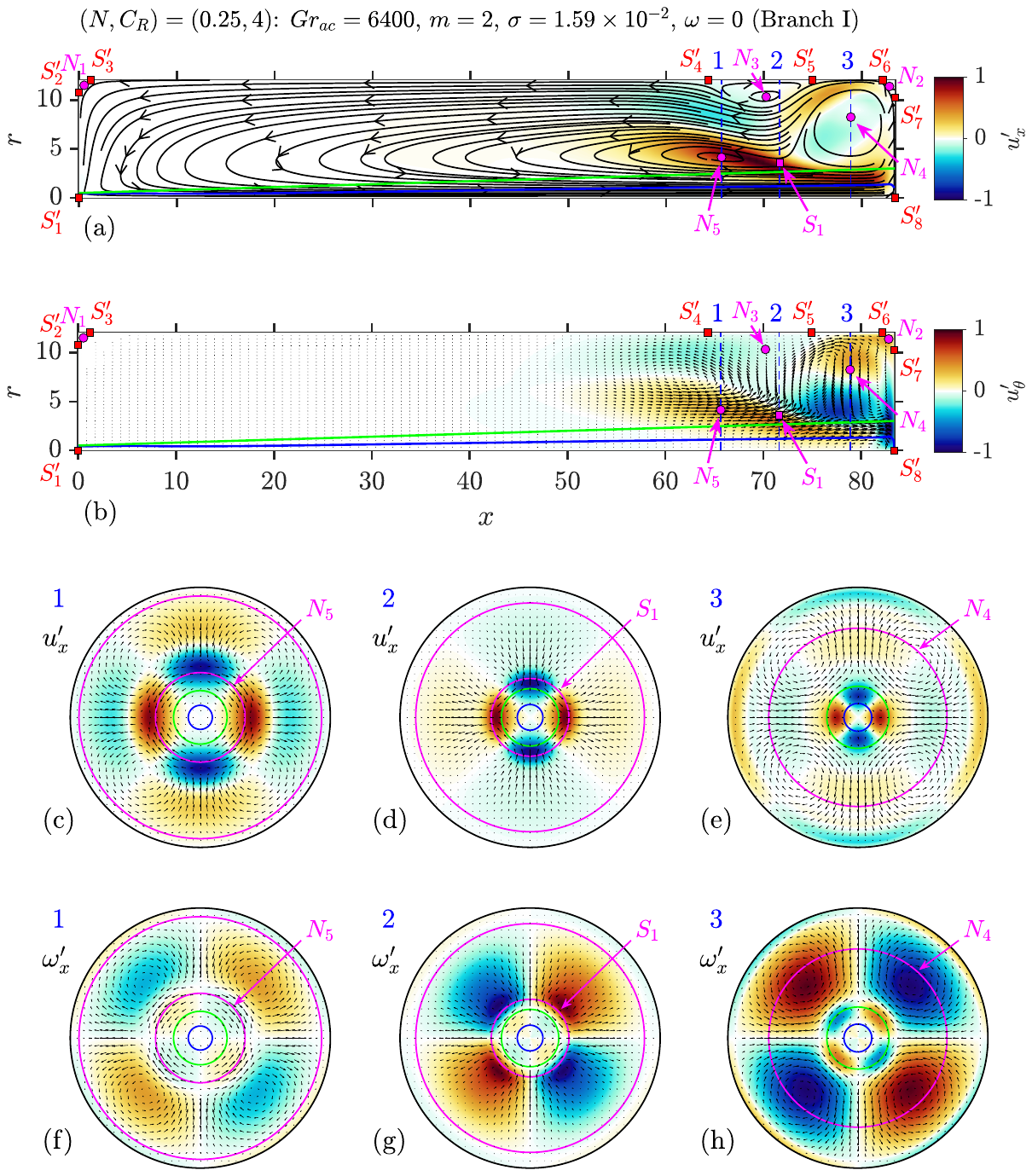}
    \caption{Leading mode for $\left( N, C_R \right)=\left( 0.25, 4 \right)$ at $Gr_{ac} = 6400$. The mode is non-oscillatory ($\omega = 0$) and unstable ($\sigma = 1.59 \times 10^{-2} > 0$). (a) Axial velocity perturbation $u'_x$ in the $\left( x, r \right)$ plane, along with the streamlines and critical points of the base velocity $\boldsymbol{U}$. (b) Velocity perturbation $\boldsymbol{u'}$, with the background color corresponding to its azimuthal component $u'_{\theta}$. (c) to (h): Slices of $\boldsymbol{u'}$ in constant-$x$ planes located by the blue vertical lines in (a) and (b), with either $u_x'$ or the streamwise vorticity perturbation $\omega_x'$ as background colour (negative in blue, positive in red). The purple circles locate the points where $U_x = 0$. In all figures, the green solid line represents the approximate beam radius~\eqref{eq:beam_radius}, and the blue solid line locates the radius where $U_x$ is 50~\% of its on-axis value. All the figures on a given row share the same colour levels.}
    \label{fig:mode_shape_case3}
\end{figure}

The topology of leading eigenmodes from Branch I is illustrated for $\left( N, C_R \right) = \left( 0.25, 4 \right)$ in figure~\ref{fig:mode_shape_case3}, with the critical points of the base flow overlaid. From the topology in the $\left(x,r\right)$ plane (figure~\ref{fig:mode_shape_case3}~(a) and (b)), the perturbation intensity as measured by $u^\prime_x$ is strongest in a region stretching from the impingement region of the jet to the shear layer between recirculations centred on $N_4$ and $N_5$. Radially, the perturbation is located within the acoustic beam (the approximate acoustic beam radius $R_{beam}$ is marked by a green line) but remains outside the jet (the points where $U_x$ is 50~\% of the on-axis velocity are marked by the blue line). The perturbation then crosses the shear layer at the boundary of the jet to reach the region between the two recirculations around $N_4$ and $N_5$. Hence, the mechanism is not associated to a shear layer instability but rather coincides precisely with the strong velocity gradients in the impingement region centred on $S^\prime_8$, where the flow turns from its incoming streamwise direction to a radial one along the end wall.  The azimuthal structure is best illustrated by contours of streamwise vorticity in constant $x$ planes containing either $N_5$, $S_1$ or $N_4$ (figure~\ref{fig:mode_shape_case3} (f)-(h)) showing a breakdown into an $m=2$ modulation. These figures also clearly show that the streamwise vortices associated to these structures do not follow the shear layer (marked by the purple line corresponding to $U_x=0$) and so exclude a shear layer instability mechanism.

\begin{figure}
    \centering
    \includegraphics[trim={0, 0cm, 0, 0cm}, clip, width=0.9\textwidth]{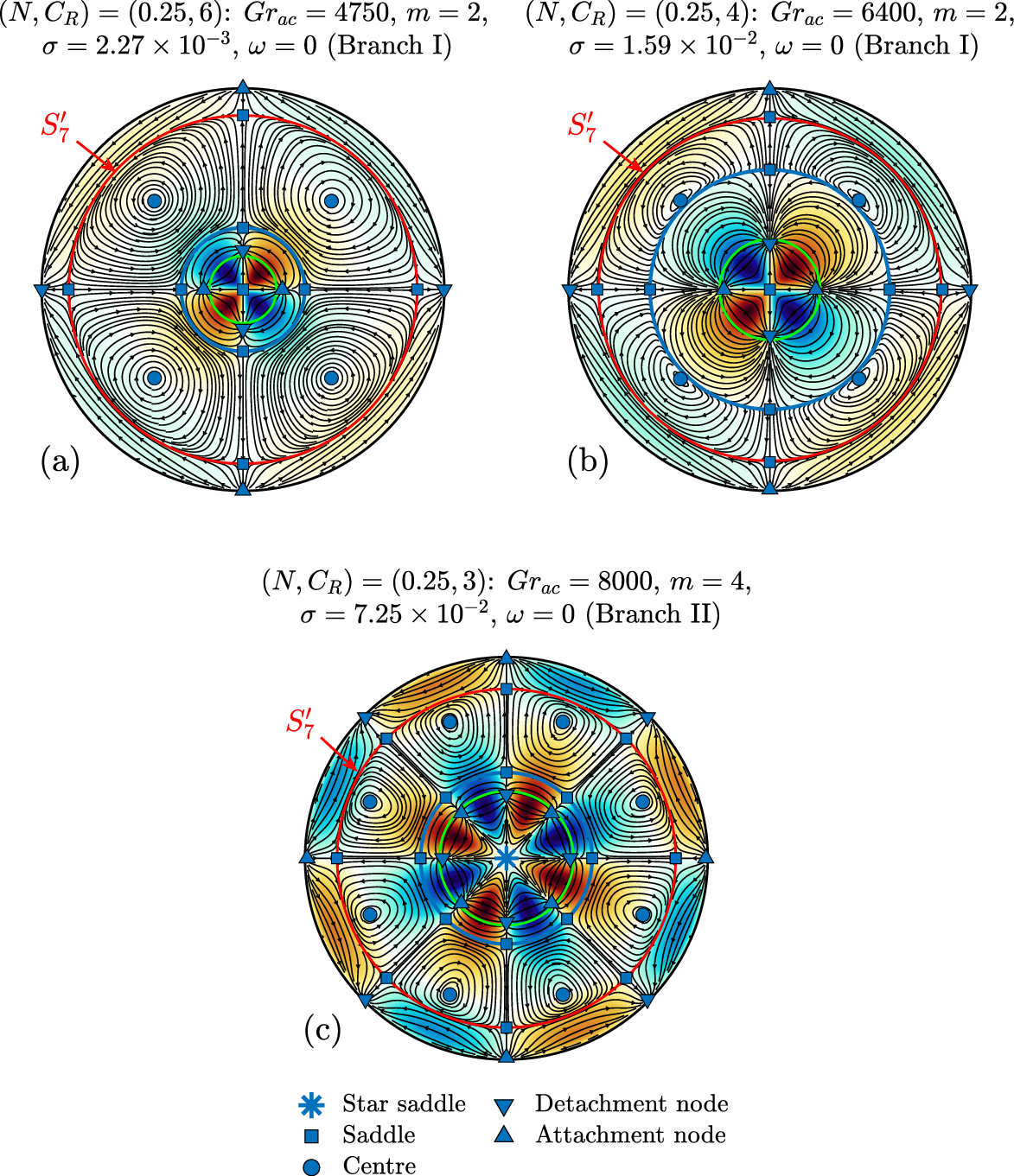}
    \caption{Lines of the skin friction perturbation stresses on the downstream wall, with the background color corresponding to the azimuthal component of the stress vector (positive in red, negative in blue). These lines are shown along with the critical points (blue symbols) at for $N = 0.25$ and for different $C_R$: (a) $C_R = 6$, (b) $4$ and (c) $3$. The approximate beam radius (equation~\eqref{eq:beam_radius}) is represented by the green circle. The base flow half saddle $S'_7$ on the downstream wall is located by the solid red circle. The solid blue circle approximates the separation line going through the saddles found at a same distance from the origin. For each case, the mode shapes are computed at forcing magnitudes $Gr_{ac}$ slightly above the instability onset.}
    \label{fig:mode_shape_skinFrictionLines_cases2_3_and_4}
\end{figure}

To best capture the differences between the modes associated to different branches of instabilities, we use the lines of the skin friction perturbation on the downstream wall, and classify their critical points. For case  $\left(N, C_R\right)= \left(0.25, 4\right)$ (figure~\ref{fig:mode_shape_skinFrictionLines_cases2_3_and_4}~(b)), 
the friction lines include the following numbers of critical points:
\begin{equation*}
\Sigma_N = 8 \, , \quad \Sigma_{N'} = 4 \, , \quad \Sigma_S = 9 \, , \quad \Sigma_{S'} = 0.
\end{equation*}
Since $\chi = 2$ for the downstream wall \citep{Foss2004}, the topological rule~\eqref{eq:topological_rule} is thus again satisfied, validating the mode structure shown in figure~\ref{fig:mode_shape_skinFrictionLines_cases2_3_and_4}~(b). The topology features attachment and detachment nodes where regions of zero azimuthal stress cause the flow to respectively attach to and separate from the wall~\citep{Delery2001}. The half-saddle of the base flow $S'_7$ in the meridional plane translates into a circular line separating a ring of Moffat corners in the outmost part of the cavity and an adjacent inner ring of four structures with closed friction lines (red line in the plots of figure~\ref{fig:mode_shape_skinFrictionLines_cases2_3_and_4}). The friction lines patterns thus feature four centres in this region (figure~\ref{fig:mode_shape_skinFrictionLines_cases2_3_and_4}~(a)). These points are located between this outer separator and an inner separator, represented by the solid blue circle. In this case the outer separator is impermeable for both the base flow and the perturbation, i.e. the perturbation field involves no mass flux between outer region of the Moffatt vortices of the base flow and the main 4-vortices structures of the perturbation. Decreasing $C_R$ narrows down the gap between the two separators, as seen by comparing cases with $C_R=6$ and $C_R=4$, respectively shown in figures~\ref{fig:mode_shape_skinFrictionLines_cases2_3_and_4}~(a) and (b). Decreasing $C_R$ reduces the space available for the friction lines to form closed loops and forms streamwise vortices that are ever more stretched in the azimutal direction.

For $C_R=3$, another local maximum of $\sigma(m)$ is first to become unstable at $Gr_{ac}^c=7752$ (figure~\ref{fig:sigma_vs_m_cases_1_to_4}~(d)). The leading modes of this new branch, which we denote as ``Branch II", correspond to non-oscillatory $m=4$ perturbations. This change of unstable branch is the first significant effect of confinement on the flow stability, and
highlights a minimum $D$ below which the leading perturbation is altered.

To understand the change of leading mode branch, we shall turn again to the patterns of the friction perturbation on the downstream wall. Here the confinement is increased to the point where the streamwise vortices between the separators of branch I are strecthed to breaking point. This happens as the centres collapse on the inner separator (blue circle), making impossible to form closed friction loops without crossing the separator. In other words, further reducing $C_R$ necessarily requires the skin friction lines patterns to be rearranged for basic topological rules to not be violated. This topological rearrangement is made possible by a different value of $m$, corresponding to a new unstable branch. The resulting Branch~II mode features a new set of saddles and nodes (figure~\ref{fig:mode_shape_skinFrictionLines_cases2_3_and_4}~(c)). The critical point at $r=0$ is a ``high-order" critical point: both the base flow axisymmetry and the $m=4$ symmetry of the mode cause the gradient of the skin friction vector to vanish, meaning that higher-order derivatives are required to classify this point. We refer to the latter as a ``star saddle". Being different in nature to classical four-branch saddles of topological weight $w_{4s}=-2$ (i.e.,  the constant by which $\Sigma_S$ is multiplied in the topological rule~\eqref{eq:topological_rule}), this type of critical point must have its own topological weight. Assuming that the topological rule~\eqref{eq:topological_rule} is satisfied, the topological weight of this 8-branch star saddle must be $w_{8s}=-6$.

\subsection{Rise of oscillatory instabilities in confined settings}

As $C_R$ is further decreased, the modes destabilising the flow become oscillatory.
\begin{figure}
    \centering
    \includegraphics[width=0.99\textwidth]{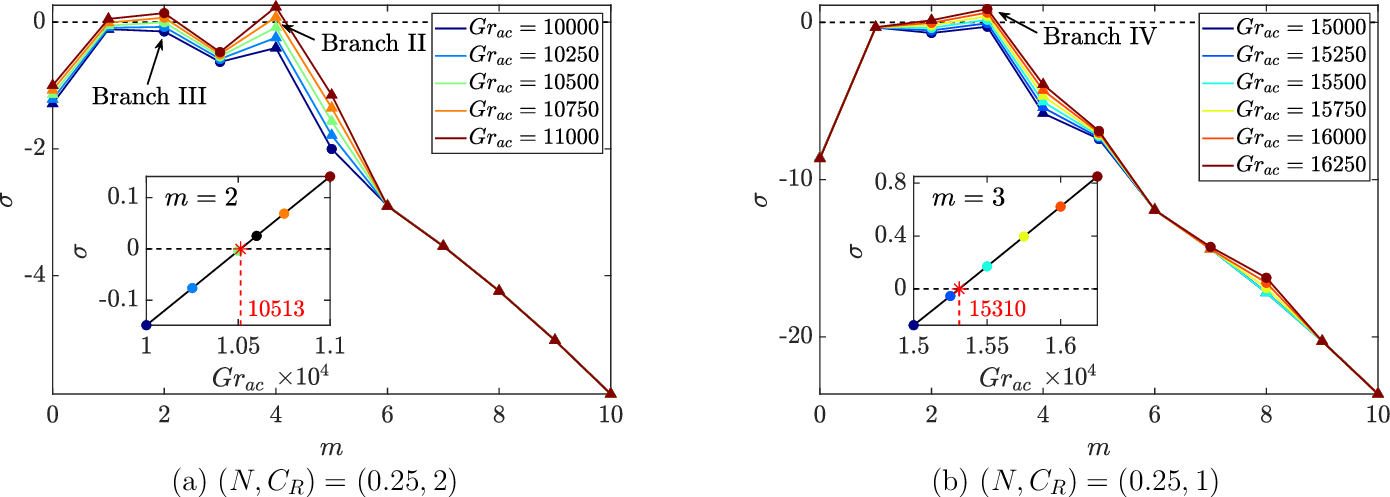}
    \caption{Growth rate $\sigma$ of the leading eigenmode as a function of the azimuthal wavenumber $m$ for (a) $\left(N, C_R\right) = \left(0.25, 2\right)$ and (b) $\left( 0.25, 1 \right)$. Oscillatory and non-oscillatory modes are represented by filled circles and triangles, respectively. The insets show the evolution of $\sigma$ with $Gr_{ac}$ for the indicated value of $m$. The critical Grashof number $Gr_{ac}^c$, obtained through spline interpolation of $\sigma$, is reported in red. The filled black symbols in the insets of (a) are additional eigenvalue computations made to improve the interpolation accuracy in the vicinity of $\sigma = 0$.}
    \label{fig:sigma_vs_m_cases_5_and_6}
\end{figure}
Indeed, the $C_R=2$ flow is destabilised at $Gr_{ac}^c = 10513$ by the growth of an oscillatory $m=2$ perturbation (figure~\ref{fig:sigma_vs_m_cases_5_and_6}~(a)). This mode is associated with a new branch that we label as ``Branch III". Interestingly, Branch~III and Branch~II coexist for this $C_R$, and the leading Branch~II mode becomes unstable at ${Gr_{ac} \approx 10630 \approx 1.01 Gr_{ac}^c}$. The $\sigma$ of the Branch II mode increases more rapidly with $Gr_{ac}$ than the $\sigma$ of the Branch~III mode, so that Branch~II becomes dominant for ${Gr_{ac} \geq 11000 \approx 1.04 Gr_{ac}^c}$.

\begin{figure}
    \centering
    \includegraphics[trim={0, 2cm, 0, 0.5cm}, clip, width=0.9\textwidth]{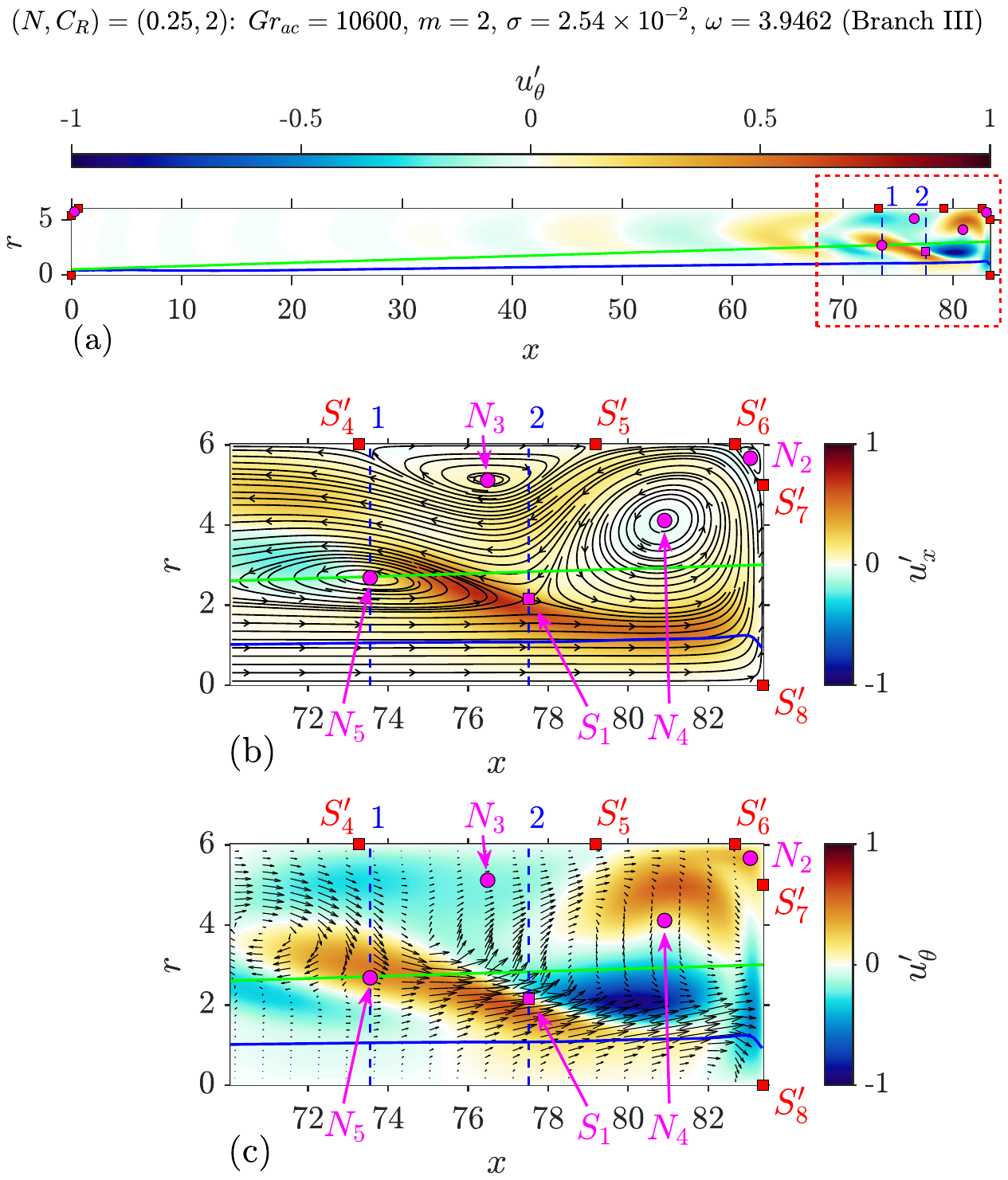}
	\caption{Leading mode for $\left( N, C_R \right)=\left( 0.25, 2 \right)$ at $Gr_{ac} =10600$. The mode is oscillatory ($\omega \neq 0$) and unstable ($\sigma = 2.54 \times 10^{-2} > 0$). (a) Azimuthal velocity perturbation $u'_{\theta}$ in the $\left( x, r \right)$ plane, along with the critical points of the base velocity $\boldsymbol{U}$. (b) Detailed view near the impingement (region framed in red in (a)) of the axial velocity perturbation $u'_x$ together with the base flow streamlines. (c) Details $\boldsymbol{u'}$ near the impingement, with $u'_{\theta}$ taken as background colour.  In all plots, the vertical dashed blues lines locate the slices shown in figure~\ref{fig:mode_shape_case5_slices}. The green solid line represents the approximate beam radius~\eqref{eq:beam_radius}, and the blue solid line locates the radius where $U_x$ is 50~\% of its on-axis value.}
	\label{fig:mode_shape_case5}
\end{figure}
\begin{figure}
	\centering
	\includegraphics[width=0.75\textwidth]{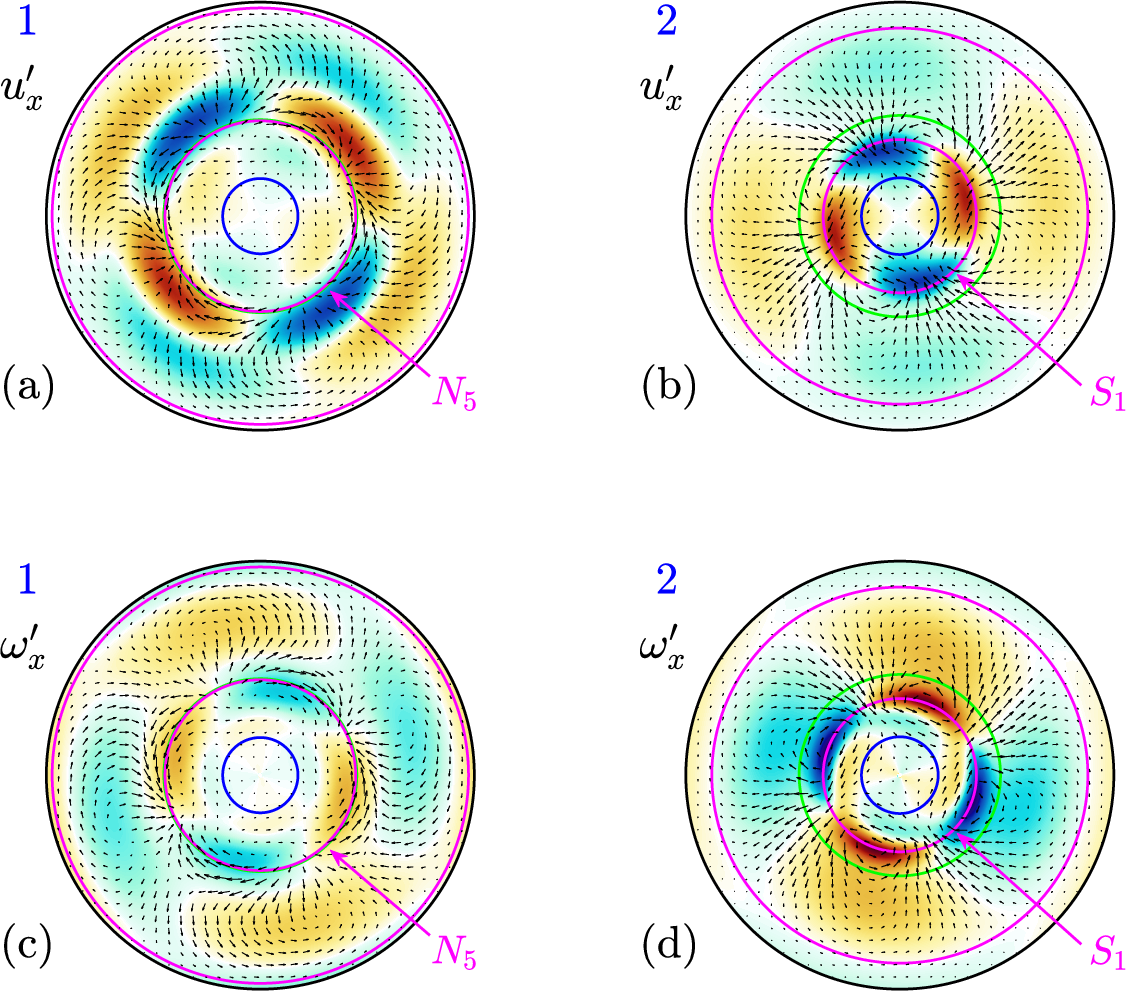}
	\caption{Leading velocity perturbation $\boldsymbol{u'}$ for $\left( N, C_R \right)=\left( 0.25, 2 \right)$ at $Gr_{ac} =10600$. The mode is displayed in constant-$x$ planes located by the vertical dashed blue lines shown in figure~\ref{fig:mode_shape_case5} and going through the base flow node $N_5$ (slice 1, left) and saddle $S_1$ (slice 2, right). Either the streamwise velocity perturbation $u_x'$ ((a)-(b)) or the streamwise vorticity perturbation $\omega_x'$ ((c)-(d)) is used as background colour (negative in blue, positive in red). In all figures, the purple circles locate the points where $U_x = 0$, the green circle represents the approximate beam radius~\eqref{eq:beam_radius}, and the blue circle locates the radius where $U_x$ is 50~\% of its on-axis value. Figures on a given row share the same colour levels.}
	\label{fig:mode_shape_case5_slices}
\end{figure}

For $C_R=2$, the increased flow confinement and the greater $Gr_{ac}^c$ cause the flow structures resulting from the jet impingement to shrink down and to be further concentrated near the downstream wall (figure~\ref{fig:mode_shape_case5}). In particular, $N_5$ and $S_1$ now lie inside the main forcing region; that was not the case for larger $C_R$. Therefore, part of the return flow near $N_5$ and $S_1$ now faces the acoustic forcing.

The velocity perturbation field displays distinct features compared to the previous ${C_R \geq 3}$ cases. First, $\vert u'_{\theta} \vert$ is no longer maximum at the impingement, but near $S_1$ instead (figure~\ref{fig:mode_shape_case5}~(c)). The flow is thus no longer destabilised by the jet impingement, but by the shear layer $N_5$ and $S_1$ instead. This is confirmed by the shape of $\boldsymbol{u'}$ and the contours of $\omega_x$ at $N_5$ and $S_1$ (figure~\ref{fig:mode_shape_case5_slices}): as for $C_R \geq 3$, the perturbation is still clustered near the shear layer around the jet where $U_x \simeq 0$. Unlike at lower confinement, the instability extends along a surface closely following this layer. Second, $u'_{\theta}$ displays a wave-like structure in the $x$-direction, the amplitude of which being significantly reduced as the distance from the downstream wall increases. This is consistent with the findings of \citet{Henry2022}, who reported similar unstable waves for a plane 2D streaming flow forced by a non-diffracting and non-attenuated beam of square cross section. These features, reminiscent of a Kelvin-Helmoltz instability, suggest that a high confinement the instability originates in the shear layer, rather than at the impingement.
\begin{figure}
    \centering
    \includegraphics[trim={0, 0cm, 0, 0cm}, clip, width=0.9\textwidth]{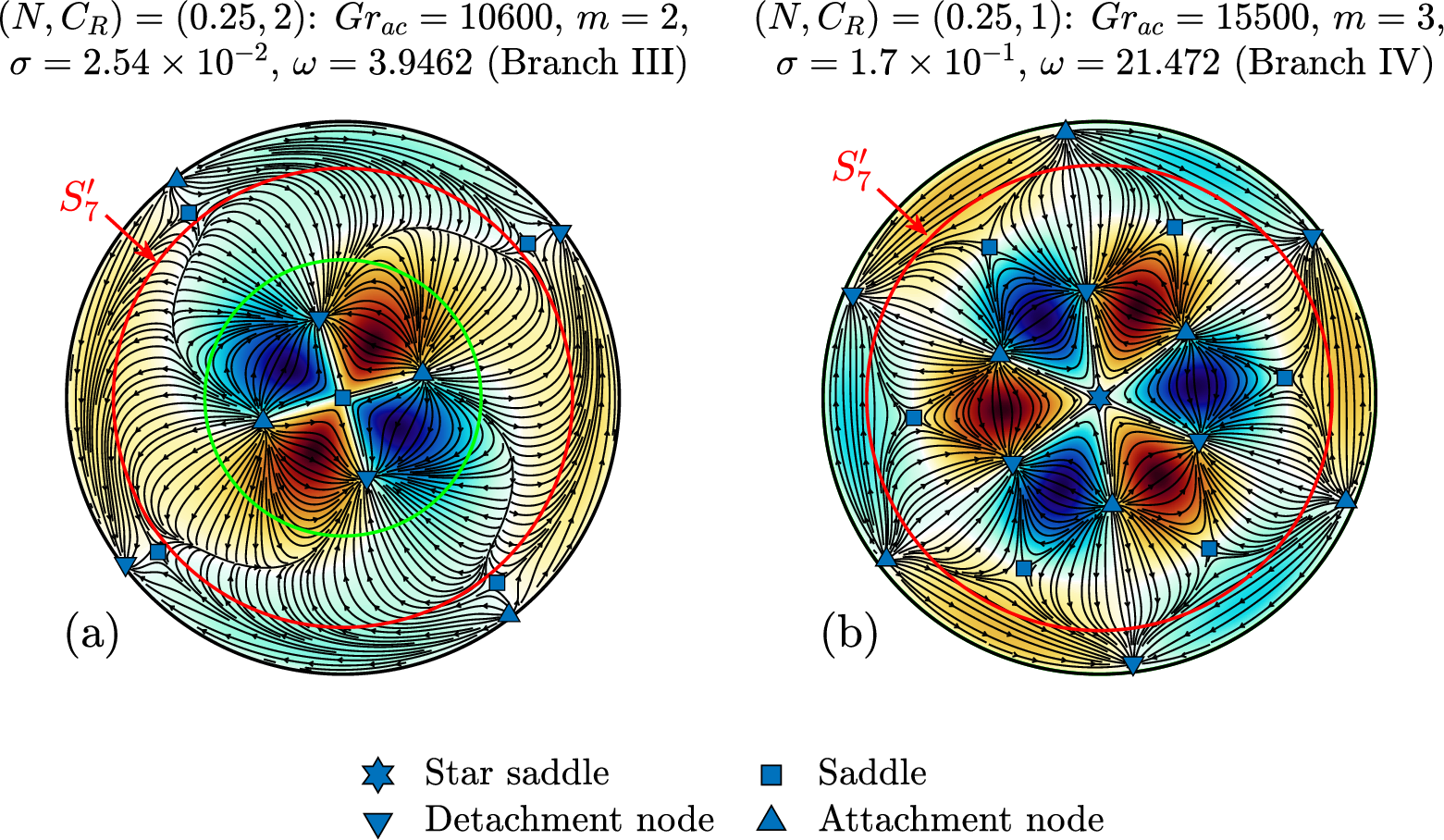}
    	\caption{Lines of skin friction stresses exerted by the leading perturbation on the downstream wall for (a) $\left( N, C_R \right) = \left( 0.25, 2 \right)$ and (b) $\left( N, C_R \right) = \left( 0.25, 1 \right)$. In each plot, the background colour represents the azimuthal component of the stress vector (positive in red, negative in blue). The friction lines are shown along with the zero mode friction critical points (blue symbols) and the approximate beam size (green circle, as defined by equation~\eqref{eq:beam_radius}). The base flow half saddle $S'_7$ is also reported in red.}
    \label{fig:mode_shape_skinFrictionLines_cases5_and_6}
\end{figure}
Finally, for $C_R = 2$, the saddles of the skin friction perturbation on the downstream wall no longer coincide with the base flow half saddle $S'_7$ (figure~\ref{fig:mode_shape_skinFrictionLines_cases5_and_6}~(a)). In other words, reducing the cavity radius creates a perturbation flux between the Moffatt roll delimited by $S'_7$ and the remaining of the flow near $x = L$.

\begin{figure}
    \centering
    \includegraphics[trim={0, 2cm, 0, 1cm}, clip, width=0.85\textwidth]{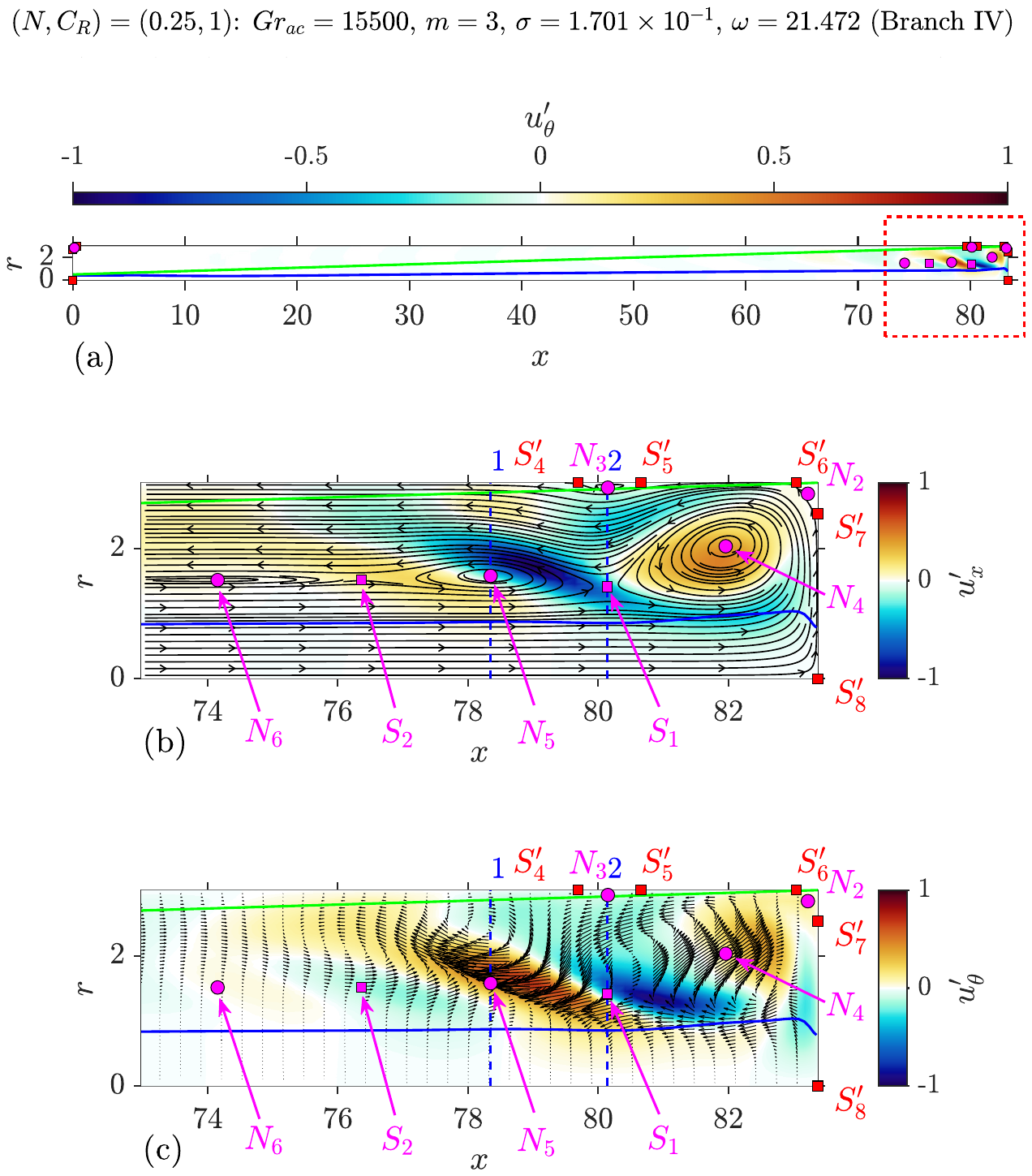}
	\caption{Leading mode for $\left( N, C_R \right)=\left( 0.25, 1 \right)$ at $Gr_{ac} = 15500$. The mode is oscillatory ($\omega \neq 0$) and unstable ($\sigma = 1.701 \times 10^{-1} > 0$). (a) Azimuthal velocity perturbation $u'_{\theta}$ in the $\left( x, r \right)$ plane, along with the critical points of the base velocity $\boldsymbol{U}$. (b) Detailed view near the impingement (region framed in red in (a)) of the axial velocity perturbation $u'_x$ together with the base flow streamlines. (c) Details of $\boldsymbol{u'}$ near the impingement, with $u'_{\theta}$ taken as background colour. }
	\label{fig:mode_shape_case_6}
\end{figure}
\begin{figure}
	\centering
	\includegraphics[width=0.8\textwidth]{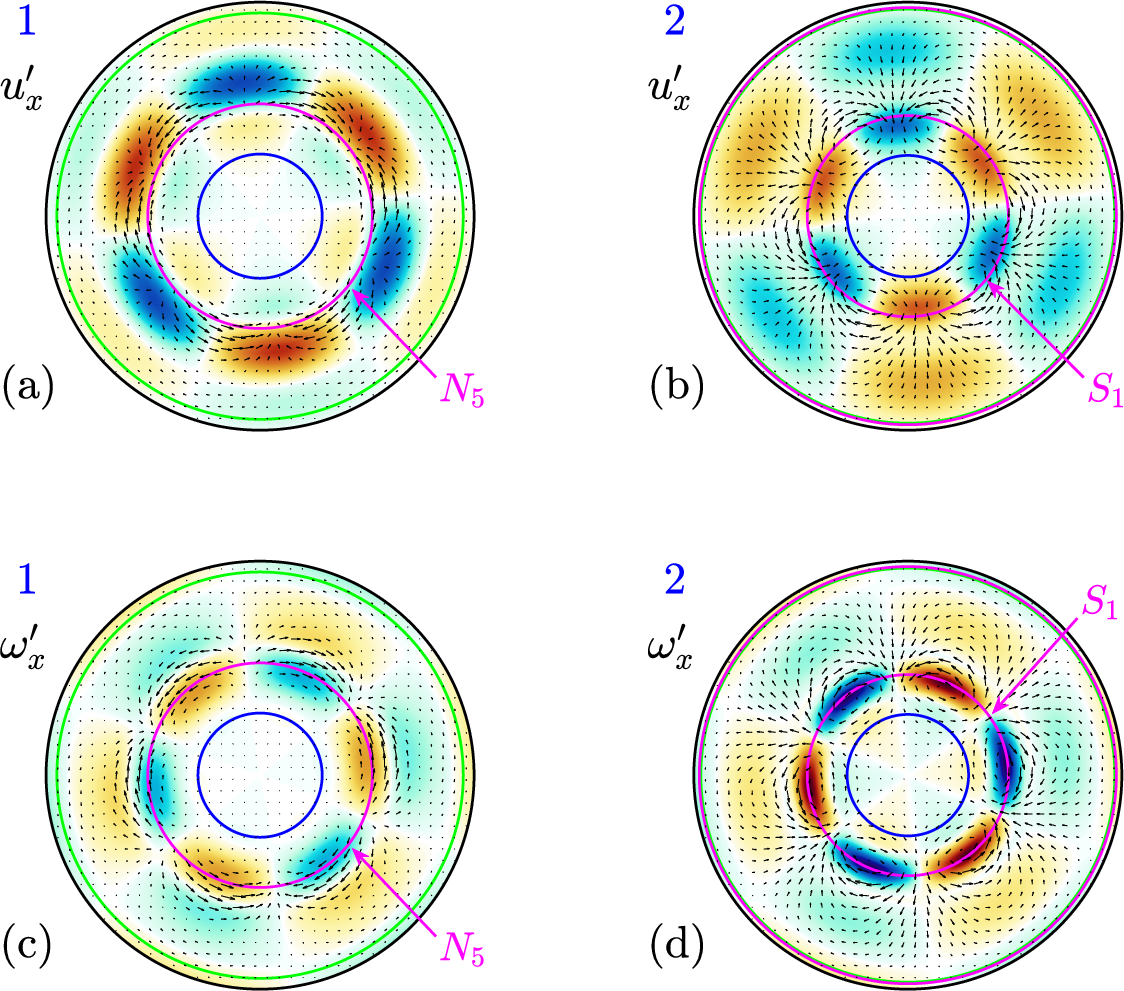}
	\caption{Leading velocity pertubation $\boldsymbol{u'}$ for $\left( N, C_R \right)=\left( 0.25, 1 \right)$ at $Gr_{ac} = 15500$. The mode is shown in constant-$x$ planes located by the blue vertical lines in figure~\ref{fig:mode_shape_case_6}, with either $u_x'$ ((a)-(b)) or the streamwise vorticity perturbation $\omega_x'$ ((c)-(d)) as background colour (negative in blue, positive in red). The purple circles locate the points where $U_x = 0$. In all figures, the green solid line represents the approximate beam radius~\eqref{eq:beam_radius}, and the blue solid line locates the radius where $U_x$ is 50~\% of its on-axis value. Figures on a given row share the same colour levels.}
	\label{fig:mode_shape_case_6_slices}
\end{figure}

For the most confined case ($C_R=1$), a new mode branch arises (figure~\ref{fig:sigma_vs_m_cases_5_and_6}~(b)). This branch, labelled as ``Branch IV", is defined by a leading $m=3$ oscillatory mode. The onset occurs at $Gr_{ac} = 15310$. At $Gr_{ac} \approx Gr_{ac}^c$, both the enhanced confinement and the greater strength of the return flow resulting from the larger $Gr_{ac}^c$ significantly affect the base flow topology near the downstream wall (figure~\ref{fig:mode_shape_case_6}). First, the size of the recirculating flow structures is greatly reduced e.g., the secondary recirculation bubble on the lateral wall nearly vanishes. Second, the reduced $C_R$ yields two new critical points $N_6$ and $S_2$ in the spatial structure of $\boldsymbol{U}$. These new critical points appear at approximately the same radial location as $N_5$ and $S_1$.

Maps of $\boldsymbol{u'}$ in the $\left( x, r \right)$ plane show that $\vert u'_x \vert$ and $\vert u'_{\theta} \vert$ are maximum near $N_5$ and $S_1$, i.e., in regions of high shear in the bulk of the base flow (figure~\ref{fig:mode_shape_case_6}~(b) and (c)). This confirms the trend observed for $C_R = 2$: confining the flow shifts the locus of the instability from the impingement to the shear layer near $N_5$ and $S_1$. This shift is also favoured by the build-up of an adverse pressure gradient slowing down the jet (inset of figure~\ref{fig:longitudinalVelocityProfiles_cases2_to_6_GrAc6400}). Besides, $\boldsymbol{u'}$ is again marginal in the jet (blue line and circles in figure~\ref{fig:mode_shape_case_6}) and creates strong $\omega_x$ at the shear layer (figure~\ref{fig:mode_shape_case_6_slices}). Finally, the leading Branch~IV mode features a ``star saddle" critical point at $r=0$ (figure~\ref{fig:mode_shape_skinFrictionLines_cases5_and_6}~(b)), as does the leading mode for $C_R=3$ (Branch~II).

To conclude, for $N \in \{ 0.25, 1 \}$ and $3 \leq C_R \leq 6$, the primary instabilities are caused by non-oscillatory perturbations originating from the jet impingement. Further confining the flow ($1 \leq C_R \leq 2$) not only relocates the origin of the instability to the shear layer between the jet and the impingement-induced recirculation structures, but also causes the leading perturbation to become oscillatory.

\section{Bifurcation characterisation}
\label{sec:characterisation_of_the_bifurcations}

\subsection{Model for the nonlinear growth and saturation of the perturbations}
\label{subsec:stuart-landau_model}

To complete the stability study, we shall now determine the criticality of the bifurcations identified with LSA. Our approach relies on the model of \citet{Landau1987} describing the growth and saturation of a perturbation near the onset of instability. This model has been extensively used to 
classify the bifurcations arising in numerous situations, such as the flow past rings \citep{Sheard2004}, the wake behind a sphere~\citep{Thompson2001} and behind one or multiple cylinders \citep{Henderson1996,Henderson1997,Carmo2008}. The Stuart-Landau model reads:
\begin{equation} \label{eq:landau_model}
    \frac{ \mathrm{d} A}{ \mathrm{d} t} = \left( \sigma + \mathrm{i} \omega \right) A - l \left( 1 + \mathrm{i} c \right) \vert A \vert ^2 A + O(A^5) \, .
\end{equation}
where $A$ is a complex perturbation, and $l$ and $c$ are two real constants. The mechanism responsible for the saturation of $A$ can be easily understood from the sign of $l$ in equation~\eqref{eq:landau_model}. For $l > 0$, the cubic term in the right-hand side of equation~\eqref{eq:landau_model} acts against the growth of the perturbation: this situation corresponds to a supercritical bifurcation. On the contrary, for $l < 0$, higher-order terms are required for $A$ to saturate: the bifurcation is thus subcritical. Therefore, only the sign of $l$ is needed to determine the nature (or criticality) of the bifurcation.

As explained in \citet{Sheard2004} and \citet{Kumar2020}, $l$ may be inferred from a more convenient form of equation~\eqref{eq:landau_model}. Plugging $A(t) = \vert A(t) \vert e^{ \mathrm{i} \phi(t) }$ into the Stuart-Landau model~\eqref{eq:landau_model} gives, after separating the real and imaginary part:
\begin{gather}
    \frac{ \mathrm{d} \log \vert A \vert }{ \mathrm{d} t} = \sigma - l \vert A \vert ^2 \, + O ( \vert A \vert^4 ), \label{eq:landau_amplitude}\\
    \frac{ \mathrm{d} \phi}{ \mathrm{d} t} = \omega - l c \vert A \vert^2 + O ( \vert A \vert^4 ) \, . \label{eq:landau_phase}
\end{gather}
The nature of the bifurcation is determined by plotting $(\mathrm{d} \log \vert A \vert / \mathrm{d} t)$ against $\vert A \vert^2$; the slope of the resulting curve is $-l$ for sufficiently small $\vert A \vert^2$, and $\sigma$ is obtained from the $y$-intercept. From this we compare $\sigma$ to its value obtained from the LSA results and so assess the accuracy of the estimate for $l$. Note that for an non-oscillatory perturbation, $\phi(t) = 0$, so only equation~\eqref{eq:landau_amplitude} is needed.

The saturated state is characterised using the steady-state forms of equations~\eqref{eq:landau_amplitude} and \eqref{eq:landau_phase}. For a supercritical ($l > 0$) bifurcation for instance, the first two terms in the right-hand side of the amplitude equation~\eqref{eq:landau_amplitude} are dominant, so that the saturated perturbation amplitude is:
\begin{equation} \label{eq:landau_saturated_amplitude}
    \vert A_{sat} \vert = \sqrt{ \frac{\sigma}{l} } \, .
\end{equation}
Furthermore, if the saturated state is a time-periodic signal of constant amplitude $\vert A_{sat} \vert$, then $\mathrm{d} \phi / \mathrm{d} t$ reduces to the saturated frequency $\omega_{sat}$. Using 
equation ~\eqref{eq:landau_saturated_amplitude}, 
the steady-state version of equation~\eqref{eq:landau_phase} gives, for the Landau constant $c$:
\begin{equation} \label{eq:landau_constant}
    c = \frac{ \omega - \omega_{sat} }{\sigma} \, ,
\end{equation}
thus allowing for all the quantities in the truncated Stuart-Landau model~\eqref{eq:landau_model} to be fully characterised.

We shall now define $A$ in terms of flow quantities. We followed \citet{Kumar2020} and based $\vert A \vert$ on a local value of a flow variable, and more particularly on $u'_{\theta}(t)$. Other choices are possible, including global measures of the flow unsteadiness \citep{Thompson2001,Sheard2004,Sherwin2005,Sapardi2017}. For oscillatory bifurcations, we considered the envelope of $u'_{\theta}(t)$. Choosing where to record $u'_{\theta}(t)$ is based on the need to obtain clean signals. To maximise the signal-to-noise ratio, we thus chose points where LSA predicted large $\vert u'_{\theta} \vert$. Depending on the values of $C_R$, these points are located near the critical points $N_5$ and $S_1$, and near the impingement. The precise locations of these points are given in table~\ref{tab:probe_locations}.

\begin{table}
    \begin{center}
        \begin{tabular}{ccc}
        $N$ & $C_R$ & $\left(x, r, \theta\right)$\\
        0.25 & 1 & $\left(82, 1.8, 0 \right)$\\
        0.25 & 2 & $\left(81, 1.3, 0\right)$\\
        0.25 & 3 & $\left(80, 2, 0\right)$\\
        0.25 & 4 & $\left(78, 6, 0\right)$\\
        0.25 & 6 & $\left(75, 2, 0\right)$\\
        1 & 6 & $\left(308, 30, 0\right)$\\
        \end{tabular}
    \end{center}
\caption{Coordinates $\left( x, r, \theta \right)$ of the points where the time series of the azimuthal velocity perturbation $u'_{\theta}$ are recorded in the nonlinear 3D-3C simulations.}
\label{tab:probe_locations}
\end{table}

From a numerical point of view, the nonlinear 3D-3C unsteady simulations were run for slightly supercritical regimes. The relative gap to $Gr_{ac}^c$ is defined by the criticality parameter $r_c$:
\begin{equation} \label{eq:criticality_parameter}
    r_c = \frac{Gr_{ac}}{Gr_{ac}^c} - 1 \, ,
\end{equation}
with $r_c > 0$ (respectively $r_c < 0$) corresponding to a supercritical (respectively subcritical) regimes. Finally, $(\mathrm{d} \log \vert A \vert / \mathrm{d} t)$ was computed from the recorded time series using centred finite differences.

\subsection{Nature of the bifurcations}
\label{subsec:nature_of_the_bifurcations}

\begin{figure}
    \centerline{ \includegraphics[width=0.99\textwidth]{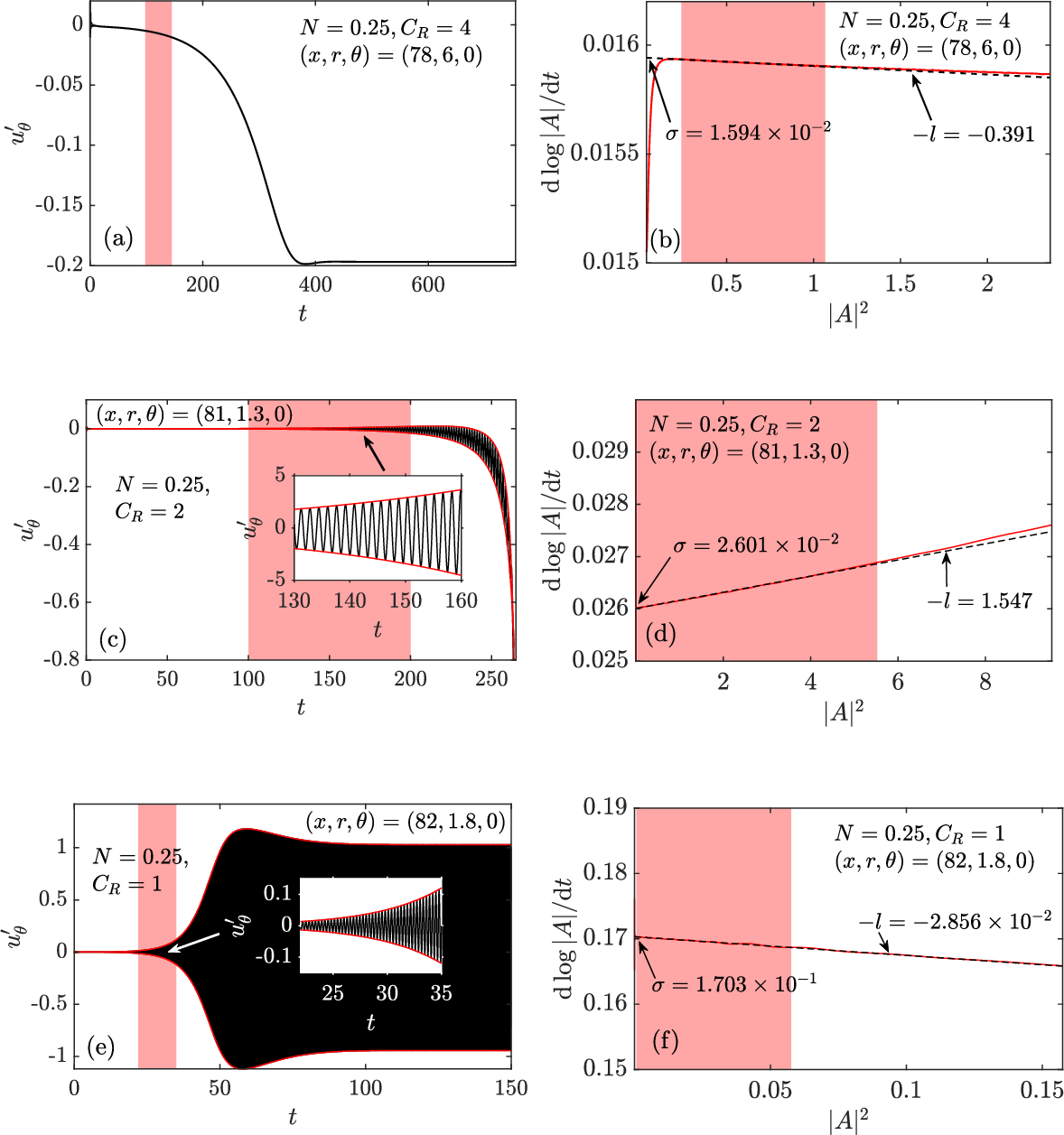} }
    \caption{Stuart-Landau analysis for (a)-(b) $\left(N = 0.25, C_R = 4\right)$ at $Gr_{ac} = 6400$ ($r_c = 0.0262$), (c)-(d) $\left(N = 0.25, C_R = 2\right)$ at $Gr_{ac} = 10600$ ($r_c = 0.0083$) and (e)-(f) $\left(N = 0.25, C_R = 1\right)$ at $Gr_{ac} = 15500$ ($r_c = 0.0124$). Left: growth and saturation of a white-noise-triggered perturbation injected at $t=0$, monitored by the time series of the azimuthal velocity perturbation $u_{\theta}'$. Right: evolution of $\mathrm{d} \left( \log \vert A \vert \right) / \mathrm{d}t$ with $\vert A \vert^2$, where $\vert A \vert$ is the amplitude of the perturbation plotted on the left figures (for the $C_R = 2$ and $1$ cases, $\vert A \vert$ is based on the envelope of $u_{\theta}'$). The growth rate $\sigma$ and the Landau coefficient $l$ are obtained by fitting equation~\eqref{eq:landau_amplitude} to the red curve. For each case, the fitting range is highlighted in red.}
    \label{fig:stuartLandau_cases_3_5_6}
\end{figure}

The Stuart-Landau analysis has been carried for all the setups listed in table~\ref{tab:parameter_values}; the cases defined by $N=0.25$ and $C_R \in \left\{ 1, 2, 4 \right\}$ are shown in figure~\ref{fig:stuartLandau_cases_3_5_6} as examples.
For each case, the initial instants of $\vert A \vert$ are dominated by noise (figure~\ref{fig:stuartLandau_cases_3_5_6}~(a), (c) and (e)). The Stuart-Landau equation~\eqref{eq:landau_amplitude} for $\vert A \vert$ is then fitted to the time series over a time interval free of the initial noise; that range is highlighted in red in all plots of figure~\ref{fig:stuartLandau_cases_3_5_6}.

For $4 \leq C_R \leq 6$, $l > 0$ near the instability onset. The leading Branch~I mode thus destabilises the flow through a supercritical circular pitchfork bifurcation~\citep{Touihri1999}. Further confining the flow yields $l < 0$ for $C_R=3$, indicating that an unstable Branch~II mode triggers a subcritical circular pitchfork bifurcation. Although the leading mode becomes oscillatory for $C_R=2$, $l$ is negative (figure~\ref{fig:stuartLandau_cases_3_5_6}~(d)): the associated Hopf bifurcation is thus subcritical. Finally, $l > 0$ for $C_R=1$ (figure~\ref{fig:stuartLandau_cases_3_5_6}~(f)), hence the unstable Branch~IV mode is associated with a supercritical Hopf bifurcation. For all cases, the difference between the $\sigma$ obtained with the nonlinear 3D-3C simulations and LSA is at most 6~\% (table~\ref{tab:stuartLandau_recap}). This excellent agreement (i) validates the LSA, and (ii) adds further confidence on the validity of the Stuart-Landau analyses.

Finding subcritical bifurcations for $C_R \in \left\{ 2, 3 \right\}$ raises the question of whether the corresponding instabilities can be triggered for $r_c < 0$. For both cases, introducing white noise for $- 0.05 \leq r_c \leq -0.01$ did not cause the flow to deviate from its axisymmetric base state. Triggering these bifurcations at $Gr_{ac} < Gr_{ac}^c$ would require different tools, e.g. the analysis of the transient growth of non-modal perturbations~\citep{Schmid2007}. Finding the minimum seeds to trigger these subcritical bifurcations is nevertheless out of the scope of the present work and shall be the subject of a dedicated study.

\begin{table}
    \begin{center}
        \begin{tabular}{ccccccccccc}
        $N$ & $C_R$ & $Gr_{ac}$ & $r_c$ & \thead{ $\sigma$ \\(LSA) } & \thead{$\sigma$ \\(nonlinear\\3D-3C)} & \thead{$\varepsilon_{\sigma}$ \\(\%)} & \thead{ $\omega$ \\(LSA)} & \thead{$\omega$ \\(nonlinear\\3D-3C)} & \thead{ $\varepsilon_{\omega}$ \\(\%) } & $\omega$ - $\omega_{sat}$ \\
        0.25 & 1 & 15500 & 0.0124 & $1.701 \times 10^{-1}$ & $1.703 \times 10^{-1}$ & 0.12 & 21.472 & 21.409 & 0.29 & 0\\
        0.25 & 2 & 10600 & 0.0083 & $2.540 \times 10^{-2}$ & $2.601 \times 10^{-2}$ & 2.35 & ~~3.946 & ~~3.958 & 0.30 & $1.166$\\
        0.25 & 3 & 8000 & 0.0320 & $7.245 \times 10^{-2}$ & $7.690 \times 10^{-2}$ & 5.79 & 0 & 0 & $-$ & $-$\\
        0.25 & 4 & 6400 & 0.0262 & $1.591 \times 10^{-2}$ & $1.594 \times 10^{-2}$ & 0.19 & 0 & 0 & $-$ & $-$\\
        0.25 & 6 & 5000 & 0.0635 & $1.384 \times 10^{-2}$ & $1.383 \times 10^{-2}$ & 0.07 & 0 & 0 & $-$ & $-$\\
        1 & 6 & 2500 & 0.1411 & $2.755 \times 10^{-3}$ & $2.896 \times 10^{-3}$ & 4.87 & 0 & 0 & $-$ & $-$\\
        \end{tabular}
    \end{center}
\caption{Comparison between the growth rates $\sigma$ and frequencies $\omega$ returned by linear stability analysis (LSA) with those obtained from nonlinear 3D-3C unsteady simulations of supercritical regimes ($r_c > 0$, see equation~\eqref{eq:criticality_parameter}). The relative errors between the values of $\sigma$ and $\omega$ obtained by LSA and the nonlinear 3D-3C simulations are given by $\varepsilon_{\sigma}$ and $\varepsilon_{\omega}$, respectively.}
\label{tab:stuartLandau_recap}
\end{table}

\subsection{Saturated states and secondary instabilities}
\label{subsec:saturated_states}

We shall finally characterise the saturated states following the perturbation growth at $r_c > 0$.

\begin{figure}
    \centering
    \includegraphics[width=0.6\textwidth]{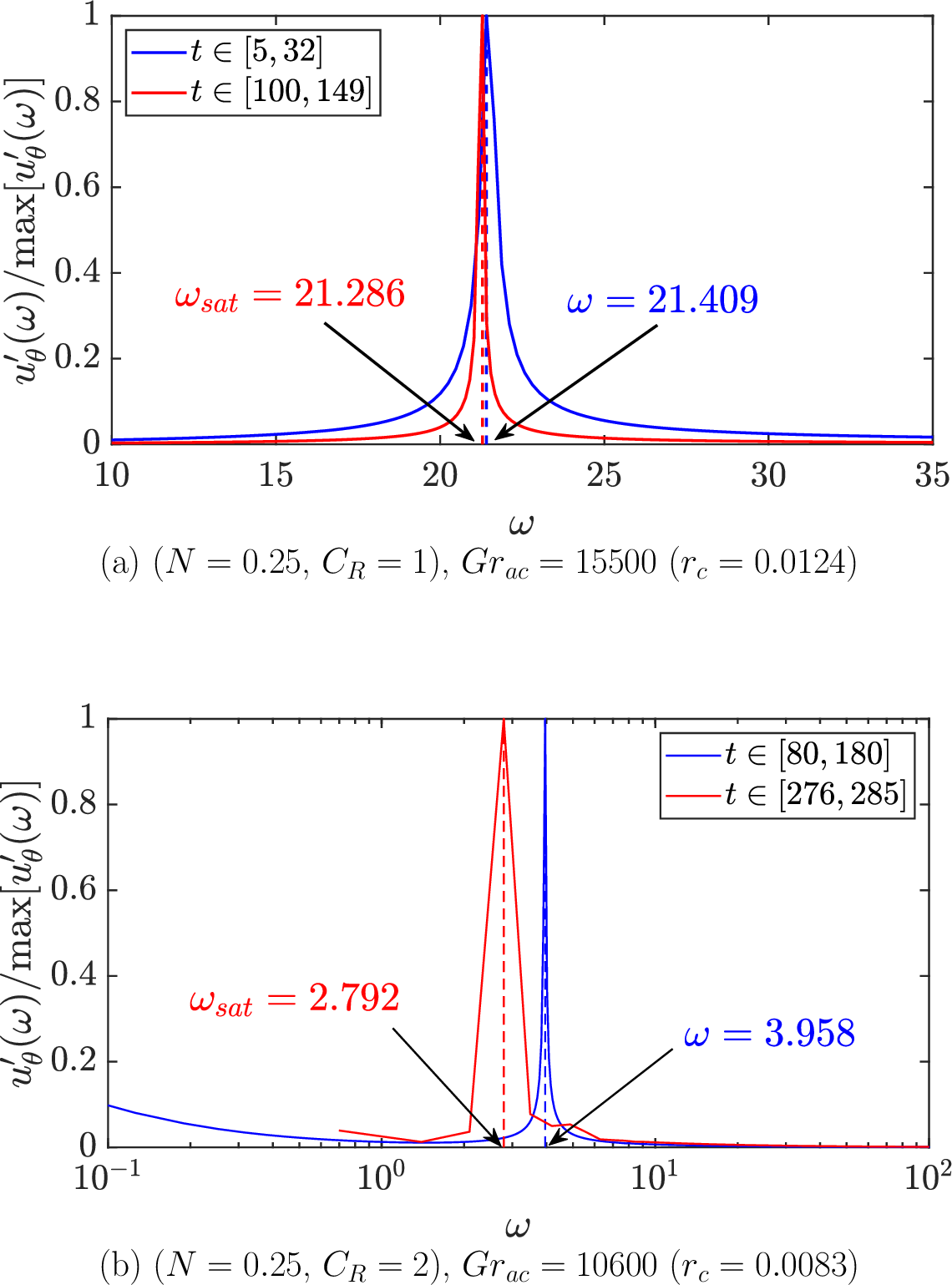}
    \caption{Frequencies obtained from the time-series of the azimuthal velocity perturbation $u'_{\theta}(t)$ for (a) $\left( N, C_R \right) = \left( 0.25, 1 \right)$ and (b) $\left( 0.25, 2 \right)$. The corresponding time-series are those displayed in figure~\ref{fig:stuartLandau_cases_3_5_6}~(e) and (c), respectively. The frequencies are computed from FFTs made in the linear regime ($\omega$, blue) and in the saturated regime ($\omega_{sat}$, red).}
    \label{fig:frequencies_cases5_6}
\end{figure}

For Hopf bifurcations, the frequency $\omega_{sat}$ of the saturated perturbation may differ from $\omega$ in the linear regime (regime of exponential perturbation growth), yielding a non-zero $c$ (equation~\eqref{eq:landau_constant}). A change of frequency is captured by taking the Fast Fourier Transform (FFT) of $u'_{\theta}(t)$ in both the linear and the saturated regimes (figure~\ref{fig:frequencies_cases5_6}).

For $\left( N, C_R\right) = \left(0.25, 2\right)$, $\omega_{sat}$ differs by 41.8~\% from $\omega$ (figure~\ref{fig:frequencies_cases5_6}~(b)). In that case, ${c = 44.83 \pm 14.63}$, and the discrepancy between the values of $\omega$ predicted by LSA and the nonlinear 3D-3C simulations is 0.30~\% (table~\ref{tab:stuartLandau_recap}). For $\left(N, C_R\right) = \left(0.25, 1\right)$, the $\omega$ and $\omega_{sat}$ obtained with the nonlinear 3D-3C simulations in the linear and saturated regimes, respectively, are almost identical (figure~\ref{fig:frequencies_cases5_6}~(a)). They also barely differ from the frequency predicted by LSA: the observed discrepancies are of 0.29~\% (table~\ref{tab:stuartLandau_recap}) and 0.87~\% in the linear and saturated regimes, respectively. Given these small differences, $\omega$ and $\omega_{sat}$ are considered to be equal, hence $c = 0$.
\begin{table}
    \begin{center}
        \begin{tabular}{cccccccc}
        $N$ & $C_R$ & $Gr_{ac}$ & $r_c$ & $l$ & $c$ & Branch & \thead{Bifurcation type} \\
        0.25 & 1 & 15500 & 0.0124 & $\left( 2.856 \pm 0.012 \right) \times 10^{-2}$ & $0$ & IV & Supercritical Hopf\\
        0.25 & 2 & 10600 & 0.0083 & $-1.547 \pm 0.001 $ & $44.83 \pm 14.63$ & III & Subcritical Hopf\\
        0.25 & 3 & 8000 & 0.0320 & $-0.794 \pm 0.003$ & $0$ & II & \thead{Subcritical circular\\pitchfork} \\
        0.25 & 4 & 6400 & 0.0262 & ~~$0.391 \pm 0.004 $ & $0$ & I & \thead{Supercritical circular\\pitchfork} \\
        0.25 & 6 & 5000 & 0.0635 & $0.142 \pm 0.001$ & $0$ & I & \thead{Supercritical circular\\pitchfork} \\
        1 & 6 & 2500 & 0.1411 & $\left( 1.971 \pm 0.006 \right) \times 10^{-2}$ & $0$ & I & \thead{Supercritical circular\\pitchfork} \\
        \end{tabular}
    \end{center}
\caption{Summary of the Stuart-Landau analysis based on local measurements of $u'_{\theta}$; these measurements are obtained from nonlinear 3D-3C unsteady simulations of slightly supercritical regimes ($r_c > 0$, see equation~\eqref{eq:criticality_parameter}). The $l$ estimates are bounded by their 95~\% confidence interval. The confidence intervals for $c$ further involve the resolution of the frequency spectra from which $\omega$ and $\omega_{sat}$ are determined. See table~\ref{tab:probe_locations} for the coordinates of the points where $u'_{\theta}$ is recorded.}
\label{tab:bifurcations_criticalities_recap}
\end{table}

For $\left( N, C_R\right) = \left(0.25, 2\right)$, the large uncertainty on $c$ is due to the rather low resolution of the saturated regime spectrum (figure~\ref{fig:frequencies_cases5_6}~(a)).
\begin{figure}
    \centerline{ \includegraphics[width=0.6\textwidth]{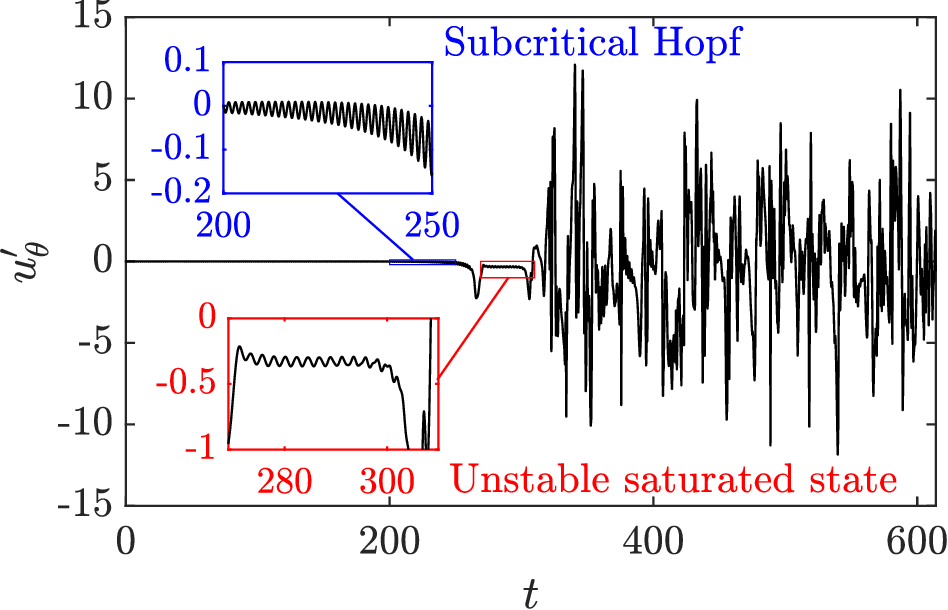} }
    \caption{Time-series of the azimuthal velocity perturbation $u_{\theta}'$ for $\left( N, C_R \right)=\left( 0.25, 2 \right)$ and ${Gr_{ac} = 10600}$ (${r_c = 0.0083}$). The signal is recorded at ${\left(x, r, \theta \right) = \left( 81, 1.3, 0 \right)}$.}
    \label{fig:primarySecondaryInstabilities_case5}
\end{figure}
That low resolution results from the short duration of the saturated regime: the latter indeed lasts approximately 10 oscillation periods after which $u'_{\theta}$ changes abruptly (figure~\ref{fig:primarySecondaryInstabilities_case5}). The saturated state is thus itself unstable and causes the flow to transition towards a highly unsteady and chaotic flow regime. Unfortunately, that secondary instability develops too abruptly to be characterised by means of a Stuart-Landau analysis.

\begin{figure}
    \centering
    \includegraphics[width=0.99\textwidth]{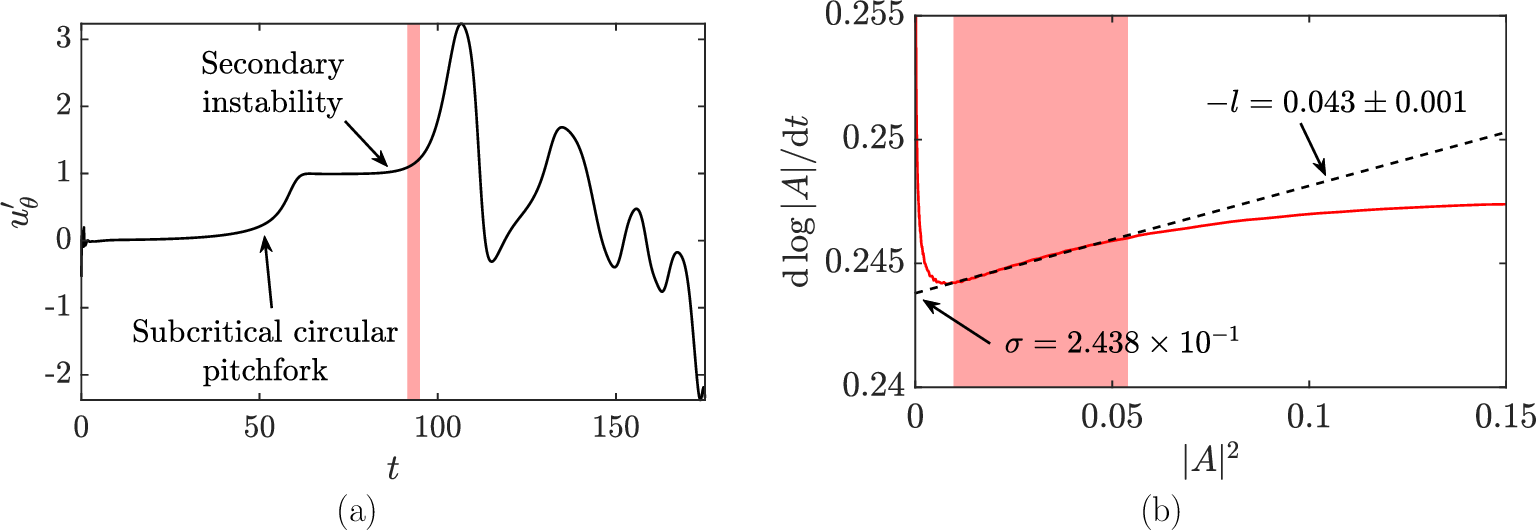}
    \caption{Stuart-Landau analysis for $\left( N, C_R \right)=\left( 0.25, 3 \right)$ at $Gr_{ac} = 8000$ (${r_c=0.0320}$). (a) Temporal evolution of the azimuthal velocity perturbation $u_{\theta}'$ obtained from nonlinear 3D-3C unsteady simulations and recorded at $\left( x, r, \theta \right) = \left( 80, 2, 0 \right)$. The perturbation originates from white noise injected at $t=0$. (b) Evolution of $\mathrm{d} \left( \log \vert A \vert \right) / \mathrm{d}t$ with $\vert A \vert^2$ (red curve), where $\vert A \vert = \vert u'_{\theta} \vert$ and is extracted from (a). The black dashed line is obtained by fitting the Stuart-Landau equation~\eqref{eq:landau_amplitude} to the red curve over the range highlighted in red. The positive slope indicates a subcritical transition.}
    \label{fig:ASjetStability_primarySecondaryInstabilities_case4}
\end{figure}

A secondary instability was also observed for $\left(N, C_R\right) = \left(0.25, 3\right)$ at $r_c = 0.0320$ (figure~\ref{fig:ASjetStability_primarySecondaryInstabilities_case4}~(a)). For that case, the saturated state is unstable to a non-oscillatory perturbation. Contrarily to the $\left(N, C_R\right) = \left(0.25, 2\right)$ case, the slow deviation from the saturated state allows for a Stuart-Landau analysis to be made (figure~\ref{fig:ASjetStability_primarySecondaryInstabilities_case4}~(b)). As the primary bifurcation, the secondary instability is found to be subcritical. However, it remains unclear whether the temporal variations of $u_{\theta}'$ observed for $t \geq 110$ correspond to a transient regime before reaching a stable steady state, or instead are the beginning of a chaotic regime. Significantly longer simulations would be necessary to answer this question.

Finally, the leading oscillatory modes for $\left( N, C_R \right) = \left( 0.25, 1 \right)$ and $\left( 0.25, 2 \right)$ are defined by a pair of complex conjugate eigenvalues corresponding to counter-propagating waves in the $\boldsymbol{e_{\theta}}$-direction. Depending on the complex amplitudes of these unstable modes, their combination results in either a standing wave or a travelling wave~\citep{Clune1993}. Although these complex amplitudes cannot be obtained from LSA, they do appear in the nonlinear 3D-3C unsteady simulations.
\begin{figure}
	\centering
	\includegraphics[width=0.72\textwidth]{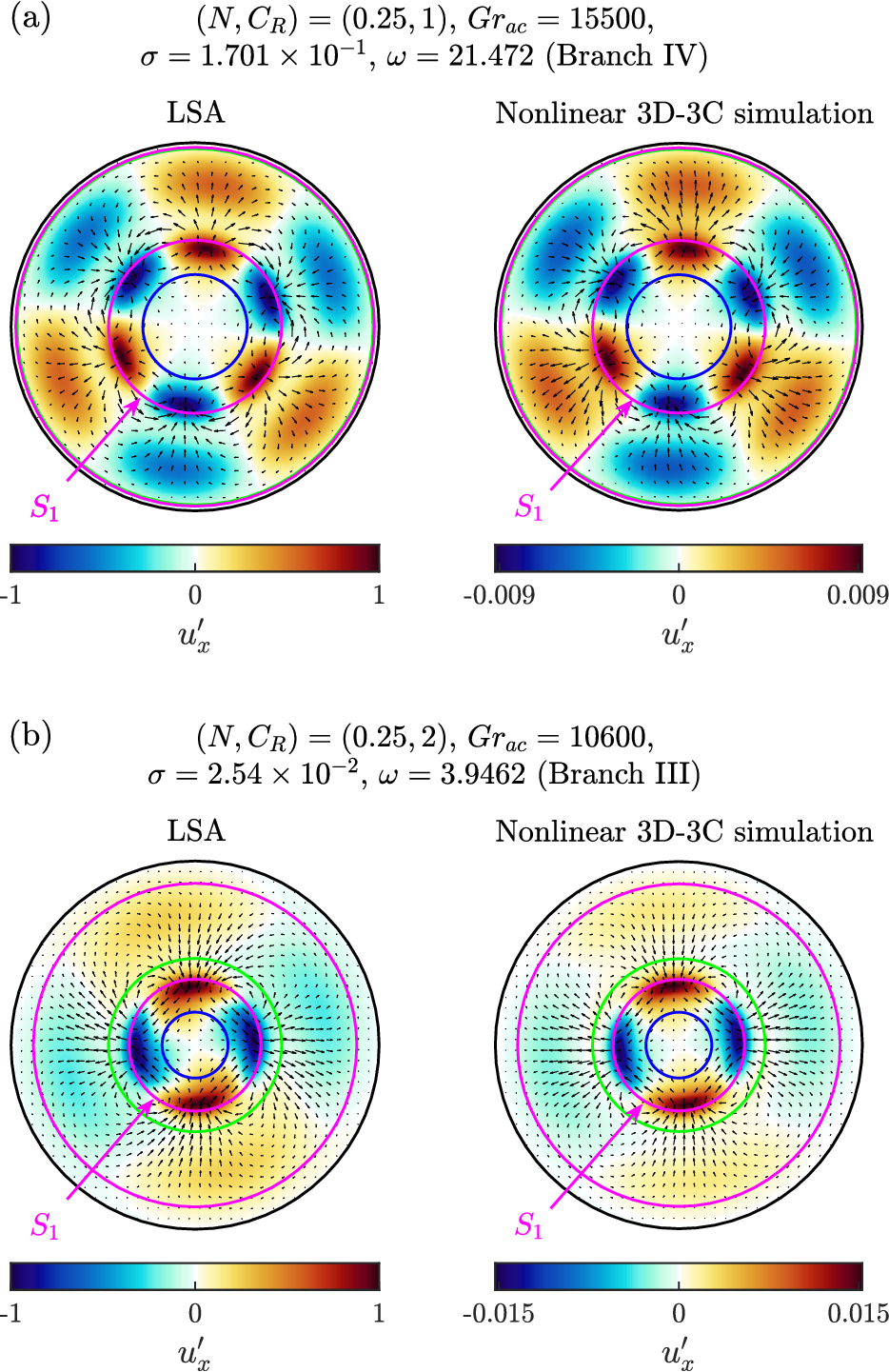}
	\caption{Comparison between the leading velocity perturbation $\boldsymbol{u'}$ obtained with LSA (left) and the perturbation obtained with nonlinear 3D-3C unsteady simulations (right) for (a) $\left( N, C_R \right) = \left( 0.25, 1 \right)$ and (b) $\left( N, C_R \right) = \left( 0.25, 2 \right)$. The results from LSA and nonlinear 3D-3C unsteady simulations are compared at the location of the axisymmetric base flow saddle $S_1$ ($x=80.15$ in (a) and $x=77.52$ in (b)). The indicated $\sigma$ and $\omega$ are those of LSA, and the nonlinear results are extracted from the early exponential growth of the perturbation. Features of the steady base velocity $\boldsymbol{U}$ are reported in both plots: the points where the sign of the streamwise velocity $U_x$ changes (purple circles), and the points where $U_x$ is equal to 50~\% of its on-axis value (blue circle). The green circle represents the approximate edge of the beam defined by equation~\eqref{eq:beam_radius}. See the supplementary movies (movies 1 and 2) for the animations of the velocity perturbations obtained from the nonlinear 3D-3C simulations.}
	\label{fig:ASjetStability_DNS_LSA_comparison_cases_5_6}
\end{figure}
The leading LSA mode shapes are compared to snapshots of the 3D-3C unsteady simulations in figure~\ref{fig:ASjetStability_DNS_LSA_comparison_cases_5_6} for $\left( N, C_R \right) = \left( 0.25, 1 \right)$ and $\left( 0.25, 2 \right)$. The snapshots are taken from the early stage of the exponential growth of the perturbation to avoid nonlinear effects on the perturbation shape. For both cases, a strong agreement in found between the Branches~III and IV mode shapes obtained by LSA and by the nonlinear 3D-3C simulations, and both perturbations correspond to waves travelling along $\boldsymbol{e}_{\theta}$ (see the supplementary movies 1 and 2).

To conclude, the thorough characterisation of the Stuart-Landau model~\eqref{eq:landau_model} for all cases (table~\ref{tab:bifurcations_criticalities_recap}) showed that the sub- or supercritical nature of the primary bifurcation is closely related to the flow confinement. For the subcritical primary bifurcations, the saturated states were found to be unstable, causing the flow to transition to an unsteady state.

\section{Conclusion}
\label{sec:conclusion}

We studied the stability of a laminar Eckart streaming jet flowing in a closed cylindrical cavity. A plane circular transducer placed at one end of the cavity radiates an axisymmetric diffracting sound beam, whose attenuation drives a steady axisymmetric jet impinging the cavity wall facing the transducer. To our knowledge, this work is the first stability analysis of a streaming jet forced by a realistic sound field that includes the effects of both diffraction and attenuation on its spatial structure. As such, this study provides a generic framework to understand the conditions in which realistic acoustic streaming jets are stable, and their path to unsteadiness when they are not. 

We assessed the effect of the cavity size on the streaming flow stability by either varying the cavity length with respect to the attenuation length ($N = 1$ and $0.25$), and by changing the ratio $C_R$ between the cavity radius and the beam radius at the downstream cavity wall ($C_R \in [ 1, 6 ]$). For each case, we (a) identified the destabilisation mechanism by analysis of the flow topology and its critical points, (b) computed the critical forcing magnitude $Gr_{ac}^c$ above which the flow is linearly unstable (Linear Stability Analysis), and (c) determined the nature of the bifurcation by means of Stuart-Landau analyses of data obtained from nonlinear 3D-3C (three dimensions, three components) unsteady simulations.
\begin{figure}
	\centering
	\includegraphics[width=0.8\textwidth]{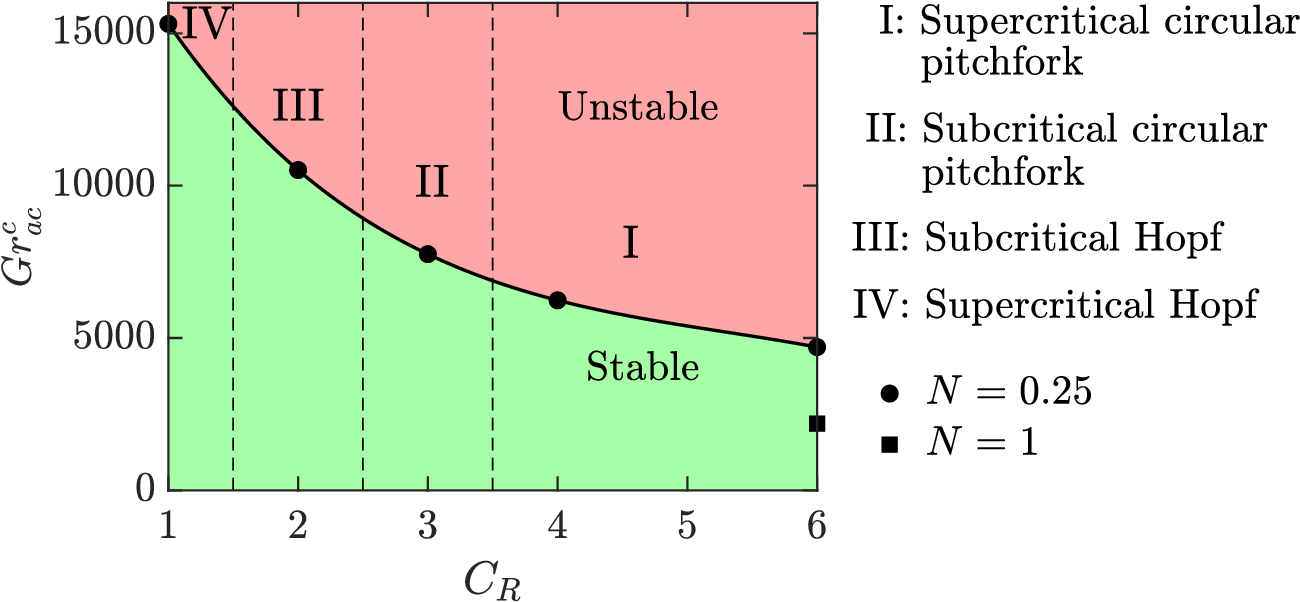}
	\caption{Variations of the critical Grashof number $Gr_{ac}^c$ with the radial confinement ratio $C_R$ defined by equation~\eqref{eq:confinement_ratio}. The $N=0.25$ cases are represented by filled discs and are interpolated by cubic splines. The $Gr_{ac}^c$ obtained for $\left(N, C_R \right)=\left(1, 6\right)$ is also reported (black square) and illustrates the destabilising effect of increasing $N$. The dashed vertical lines are arbitrarily placed in between cases having different primary bifurcation types.}
	\label{fig:evolution_critical_Grashof}
\end{figure}
The main outcomes of this work are summarised below:
\begin{enumerate}
    \item For all cavity sizes, recirculating structures originating from the jet impingement are formed near the downstream wall. The primary bifurcations result from the destabilisation of these structures, rather than from the destabilisation of the jet itself.
    \item Reducing the cavity size has a stabilising effect (figure~\ref{fig:evolution_critical_Grashof}): the critical $Gr_{ac}$ for the onset of instability ranges from $Gr_{ac}^c=2191$ for ${\left( N, C_R \right)=\left( 1, 6 \right)}$ (less-confined case) to $Gr_{ac}^c=15310$ for ${\left( N, C_R \right)=\left( 0.25, 1 \right)}$ (most confined case). For $4 \leq C_R \leq 6$, the leading perturbation is non-oscillatory (Branch~I). This unstable mode displays large perturbation velocities near the jet impingement. The first significant effect of confinement occurs for $C_R=3$: the leading mode remains non-oscillatory, but its topology changes (Branch~II). The reason for the change of topology is that the mode corresponding to Branch I with an $m=2$ azimuthal wavenumber becomes radially so confined that it becomes azimuthally stretched to the point of breaking up. At this point, an $m=4$ mode from Branch II emerges. Flows in further confined settings are destabilised by oscillatory perturbations travelling in the azimuthal direction (Branch~IV for $C_R=1$). For $C_R=2$, these perturbations also propagate backwards, i.e., towards the source (Branch~III). Overall, reducing $C_R$ (a) increases the shear in a layer separating counter-rotating structures in the bulk of the flow, and (b) enhances an adverse pressure gradient slowing down the jet. As a consequence, reducing $C_R$ moves the locus of instability from the jet impingement to the shear layer between the jet and the recirculating structures.
    \item Finally, changing the flow confinement significantly affects the nature of the primary bifurcations. Whilst the unstable modes of Branches~I and IV respectively trigger supercritical pitchfork and Hopf bifurcations, the bifurcations associated with Branch~II and III modes are subcritical.
\end{enumerate}

The subcritical nature of bifurcations for the $C_R \in \{2, 3\}$ cases raises the question of whether these bifurcations could be triggered $Gr_{ac} < Gr_{ac}^c$. The nonlinear 3D-3C unsteady simulations carried out between ${0.95 \, Gr_{ac}^c}$ and ${0.99 \, Gr_{ac}^c}$ revealed that white-noise-based perturbations could not ignite subcritical transitions towards another state. Whether this scenario could occur remains uncertain at this stage and shall require further analysis, for instance by finding the initial disturbances maximising the energy growth over a finite time range~\citep{Schmid2007}. Other examples of convective of shear flows exhibit similar subcritical bifurcations where transition away from the steady state is not easily triggered in the subcritical regime \cite{Kumar2020,Camobreco2020,Camobreco2021,Camobreco2023,Huang2024}. The stability and transition to turbulence in these flows are very different to the classical shear flows such as Couette or Poiseuille flows, where the subcritical nature of the bifurcation favours transition to turbulence via mechanisms that radically differ from those of the LSA. \citet{Camobreco2023} and \citet{Huang2024} showed that transition to turbulence in quasi-2D and 2D shear flows could be triggered only in mildly subcritical regimes and that the mode mediating the transition was in fact the leading LSA eigenmode. In acoustically-driven jets, the instability mainly originates from shear, but the base flow is significantly more complex than in these examples. Hence an interesting question for further work is whether when the transition is weakly subcritical, the leading eigenmodes underpin any transition to unsteadiness and possibly to turbulence, as they do in 2D shear flows.

For $C_R \in \left\{ 2, 3 \right\}$, nonlinear 3D-3C unsteady simulations at supercritical $Gr_{ac}$ showed that the growth and saturation of the initial white-noise perturbation were followed by the development of a secondary instability. For $C_R=2$ in particular, that secondary instability triggered a transition towards an unsteady state. This raises the question of whether this transition is the trace of an edge state, i.e., if there exists a threshold for the initial perturbation energy below which the saturated state transitions back to the initial steady axisymmetric state. Answering these questions would require further analysis.

Finally, the obtained $Gr_{ac}^c$ correspond to relatively low forcing magnitudes, as experiments usually operate at higher $Gr_{ac}$ to ease the measurement of streaming velocities~\citep{Kamakura1996,Mitome1998,Moudjed2014}. Our study thus sheds light on the rather unstable nature of acoustic streaming jets, and confirms the relevance of numerical simulations and LSA to identify the primary bifurcation of an Eckart streaming jet that is otherwise delicate to observe experimentally. This work is thus an important milestone towards the thorough understanding of the rich dynamics of streaming flows \citep{Cambonie2017,Launay2019} and their transition to turbulence. Conversely, the strong variation of $Gr_{ac}^c$ with confinement offers a convenient way to manipulate the nature of the flow forced in specific configurations: where turbulence is desired for mixing or other purposes, immersing the acoustic transducer with a plate cutting the jet in an open region of the flow would favour the instabilities conducive to turbulence. If, by contrast, a well-controlled, laminar flow is sought, the transducer could be housed at one end of a thin cylinder open towards the forcing region at the other end: such a configuration would suppress the shear instability at the edge of the jet whilst still injecting the required amount of momentum into the flow. The analysis carried out in this paper offers predictive tools that may be easily adapted to design such a device.

\backsection[Supplementary data]{\label{SupMat}Supplementary movies are available at \textcolor{red}{INSERT LINK TO SUPPLEMENTARY MOVIES}}

\backsection[Acknowledgements]{The support from the PMCS2I (Ecole Centrale de Lyon) and from the High Performance Computing Centre of the Faculty of Engineering, Environment and Computing (Coventry University) for the numerical  calculations is gratefully acknowledged. The authors would like to thank Laurent Pouilloux (Ecole Centrale de Lyon) and Alex Pedcenko (Coventry University) for their availability and for providing help at any stage of the project. The authors also wish to thank Hugh M. Blackburn for the support on the use of the spectral-element code Semtex.

\noindent For the purpose of Open Access, a CC-BY public copyright licence
has been applied by the authors to the present document and will
be applied to all subsequent versions up to the Author Accepted
Manuscript arising from this submission.}

\backsection[Funding]{This work was carried out as part of the BRASSOA project supported by the Institut Carnot Ingénierie@Lyon and by a Royal Society International Exchange grant (Ref.~IES\textbackslash R2\textbackslash 202212). AP acknowledges support from EPSRC through grant No. EP/X010937/1.}

\backsection[Declaration of interests]{The authors report no conflict of interest.}

\backsection[Author ORCIDs]{B. Vincent, https://orcid.org/0000-0002-9937-7494; A. Kumar, https://orcid.org/0000-0002-6026-5727; D. Henry, https://orcid.org/0000-0002-7231-7918; S. Miralles, https://orcid.org/0000-0002-4701-8609; V. Botton, https://orcid.org/0000-0002-0180-3089; A. Pothérat, https://orcid.org/0000-0001-8691-5241
}

\bibliographystyle{jfm}
\bibliography{references_with_DOI}

\end{document}